\documentclass[twocolumn,showpacs,amsmath,amssymb,prb,graphics,graphicx]{revtex4-1}
\usepackage{bm}
\usepackage[pdftex]{graphicx}
\usepackage{dcolumn}
\begin{document}
\title{Orbital upper critical field and its anisotropy of clean one- and two-band  superconductors }

\author{V.~G.~Kogan}
\email{kogan@ameslab.gov}
\affiliation{The Ames Laboratory and Department of Physics \& Astronomy,
   Iowa State University, Ames, IA 50011}

\author{R.~Prozorov}
\email{prozorov@ameslab.gov}
  \affiliation{The Ames Laboratory and Department of Physics \& Astronomy,
   Iowa State University, Ames, IA 50011}

\date{16 July 2012}

  \begin{abstract} The Helfand-Werthamer (HW) scheme\cite{HW}  of evaluating the orbital upper critical field is generalized to anisotropic superconductors in general, and to two-band clean materials, in particular. Our formal procedure differs from those in the literature;   it reproduces not only the isotropic HW limit,   but also the results of calculations   for the two-band superconducting MgB$_2$\cite{MMK,DS} along with the existing data on $H_{c2}(T)$ and its anisotropy $\gamma(T)=H_{c2,ab}(T)/H_{c2,c}(T)$ ($a,c$ are the principal directions of a uniaxial crystal). Using  rotational ellipsoids as model Fermi surfaces we apply the formalism developed to study $\gamma(T)$ for a few different anisotropies of the Fermi surface and of the order parameters. We find that even for a single band d-wave order parameter $\gamma(T)$ decreases on warming, however, relatively weakly. For order parameters of the form $ \Delta(k_z) =  \Delta_0(1+\eta\cos k_za)$,\cite{Xu}   according to our simulations $\gamma(T)$ may either increase or decrease on warming  even for a single band depending  on the sign of   $\eta$. Hence, the common  belief that the multi-band Fermi surface is responsible for the temperature variation of $\gamma$ is proven incorrect.

For two s-wave gaps,   $\gamma $   decreases on warming for all Fermi shapes examined.
For two   order parameters of the form $ \Delta(k_z) =  \Delta_0(1+\eta\cos k_za)$ presumably relevant for pnictides,  we obtain $\gamma(T)$ increasing on warming provided  both $\eta_1$ and $\eta_2 $ are negative, whereas for $\eta$'s\,$>0$, $\gamma(T)$ decreases.
   We  study the ratio of the two order parameters at  $H_{c2}(T)$ and find that the ratio of the small gap to the large one  does not vanish at any temperature even at  $H_{c2}(T)$, an indication that this does not happen at lower fields.

  \end{abstract}
%\pacs{74.20.-z,74.76.-w,74.50.+r,85.25.Cp}

\maketitle

\section{Introduction}

The seminal work  of Helfand and Werthamer (HW)\cite{HW} on  the temperature dependence of the upper critical field $H_{c2}(T)$ is routinely applied to analyze data on new superconductors despite the fact that HW considered the isotropic s-wave case whereas in themajority of new materials both Fermi surfaces and order parameters are strongly anisotropic. The problem of $H_{c2}(T)$   has been studied  theoretically  for   anisotropic situations as well: for layered systems,\cite{LNB} for a few cases of hexagonal anisotropy of the Fermi surface and of the order parameter,\cite{Klemm1} for the two-band MgB$_2$, \cite{Gurevich1,MMK,DS,Gurevich2,Palistrant}   for d-wave cuprates, see. e.g., Ref.\,\onlinecite{Maki} and references therein,    and in a comprehensive work  in Ref.\,\onlinecite{Kita},  to name a few.

A ubiquitous feature of $H_{c2}$ in many new   superconductors is the temperature dependent   anisotropy parameter $\gamma= H_{c2,ab }/H_{c2,c}$ (for uniaxial materials),   the property absent in conventional one-band isotropic s-wave materials. For example, $\gamma(T)$ of MgB$_2$ decreases on warming,\cite{Budko} whereas for many Fe-based  materials $\gamma(T)$ increases with increasing $T$.\cite{Budko-Canfield,Gurevich3,China}  Up to date, the $T$ dependence of the anisotropy parameter $\gamma$ is considered by many  to be caused by a multi-band character of the materials in question with the commonly given reference to the example of MgB$_2$. To our knowledge,  no explanation was offered for the ``unusual"  $\gamma(T)$ of pnictides.

In this work we develop a method to evaluate both $ H_{c2,c}(T)$ and $\gamma(T)$ that can be applied with minor modifications to various situations of different order parameter symmetries and Fermi surfaces, two bands included. Having in mind possible applications for data analysis, we  provide only  the necessary minimum of analytic developments resorting to numerical methods as soon as possible.

The upper critical field is affected by many factors: magnetic structures that may coexist or interfere with superconductivity, the paramagnetic limit, scattering, etc.  In this work we  have in mind to establish a qualitative picture of how the general features of  anisotropic Fermi surfaces and   order parameters
affect $H_{c2}$ and its anisotropy, in particular.

Since the paramagnetic effects and the possibility of Fulde-Ferrel-Larkin-Ovchinnikov phase are not included in our scheme (one can find a comprehensive discussion of these questions in Ref.\,\onlinecite{Gurevich2}), applications to   materials with high $H_{c2}(0)$ should be carried out with care; our results can prove useful for interpretation of data at temperatures where $H_{c2}(T)$ does not exceed the paramagnetic limit.

As far as applications to two-band materials are concerned, we note that our formalism applies only to superconductors with a single critical temperature $T_c$. More exotic possibilities of two component condensates with two distinct $T_c$'s are out of the scope of this paper; these are considered on general group-theoretical symmetry grounds, e.g., in Refs.\,\onlinecite{Sigrist,Joynt}.

Fine details of  Fermi surfaces are unlikely to strongly influence   $H_{c2}(T)$ because, as is well-known and shown explicitly below, only the integrals over the whole Fermi surface enter  equations for $H_{c2}(T)$.  Circumstantial evidence for a weak connection between fine particularities of the Fermi surface and  $H_{c2}(T)$ is provided by the very fact that the HW isotropic  model works so well for many one-band s-wave materials, although their Fermi surfaces  hardly resemble a sphere,  take e.g. Nb.   Another example is given by the calculations of Refs.\,\onlinecite{DS} and \onlinecite{MMK} for MgB$_2$ based on different model Fermi surfaces, but giving similar results reasonably close to  the measured $H_{c2}(T)$. We, therefore, model actual Fermi surfaces of uniaxial materials of interest here by  rotational ellipsoids  (spheroids) choosing them as to have averaged squared Fermi velocities equal to the measured values or to those calculated for realistic Fermi surfaces. This idea, in fact,  has been employed by Miranovic {\it et al.} for MgB$_2$.\cite{MMK}
  Also, we tested our method on a rotational hyperboloid as an example of open Fermi surfaces, Appendix D. This work is still in progress. We intend to study more about effects of open Fermi surfaces (or two-band combinations of closed and open surfaces) on $H_{c2}(T)$.

We focus in this work on the clean limit for two major reasons. First, commonly after discovery of a new superconductor, an effort is made to obtain as clean single crystals as possible since  those provide a better chance to study the underlying physics. Second, almost as a rule, new materials are multi-band so that the characterization of scattering leads to a multitude of scattering parameters which cannot be easily controlled or separately measured. Besides, in general, scattering suppresses the anisotropy of $H_{c2}$, the central quantity of interest in this work. We refer readers interested in effects of scattering to
 a number of successful dirty-limit microscopic models,  e.g. to Ref.\,\onlinecite{Gurevich1} and references therein.

 In the next Section, we begin  with the general discussion of the $H_{c2}$ problem for   arbitrary Fermi surfaces and order parameters. We show  that in the isotopic limit our approach is reduced to that of HW.  The basic HW approach is then applied to anisotropic situations. The derivation involves rescaling of the coordinates and, therefore, necessitates  co- and contravariant vectors representations.  In our view, disregarding this necessity
may lead to incorrect conclusions. Another formal feature of our approach should be mentioned: we avoid the minimization relative to the actual coordinate dependence of the order parameter in  the mixed state  often employed  to find   $H_{c2}(T)$.\cite{MMK,DS,Rieck,Maki,Palistrant} In this sense, our method is close to the original HW work that stresses that $H_{c2}(T)$ is actually an eigenvalue problem.

Next, we formulate the  problem for two s-wave bands and show that along with $H_{c2}(T)$ and its anisotropy one can find the ratio of order parameters on two bands, a quantity that up to date has been studied only in  zero-field. We then  formulate the problem   for two bands with order parameters of different symmetries. To show how the method actually works, we consider in detail  one or two bands with Fermi surfaces as rotational ellipsoids. The method is demonstrated on the well-studied example of MgB$_2$.

The anisotropy parameter $\gamma $ is shown to depend on temperature even for the one-band case for other than s-wave   order parameters. This dependence is weak   the d-wave materials with closed Fermi surfaces, is stronger for open Fermi shapes, and is   stronger yet  for order parameters of the form  $ \Delta =  \Delta_0(1+\eta\cos k_za)$, one of the candidates for Fe-based materials.\cite{Xu} Moreover, $\gamma(T)$ increases or decreases on warming depending on the sign of the coefficient $\eta$, in other words, on whether $\Delta$ is maximum or minimum at $k_z=0$.
These results challenge  the common belief that temperature dependence of $\gamma $ is always related to  the multi-band topology of Fermi surfaces.

For two bands,  after checking the method on MgB$_2$, we focus on situations with dominant inter-band coupling which is relevant for Fe-based materials. While in most cases we have considered, $H_{c2,c}(T)$ is qualitatively similar to the HW curve, the anisotropy parameter $\gamma(T)$ may show a non-monotonic $T$ dependence even for s-wave order parameters depending on the Fermi surface shapes and densities of states (DOS). Most interesting are the order parameters   $ \Delta =  \Delta_0(1+\eta\cos k_za)$ which yield nearly linear increase of  $\gamma(T)$ on warming, a ubiquitous feature seen in many Fe-based superconductors.

%%%%%%%
\section{The problem of $\bm{H_{c2}}$}
%%%%%%%%

Our approach is basically that of HW, although formally the equations we solve for $H_{c2}(T)$ are different and can be applied to anisotropic and multi-band situations.  To establish the link to HW, we start the discussion with the one-band  case. The problem of the 2nd order phase transition at $H_{c2}$ is addressed here on the basis of linearized (the order parameter $\Delta\to 0$) quasiclassic Eilenberger equations.\cite{Eilenb}  For clean materials we have:
\begin{eqnarray}
&&(2\omega +{\bm v}\cdot {\bm \Pi})\,f=2\Delta /\hbar\,,\label{eil1}\\
&&\frac{\Psi }{2\pi T} \ln \frac{T_{c}}{T}= \sum_{\omega
>0}^{\infty}\Big(\frac{\Psi}{\hbar\omega}-\langle \Omega \, f \rangle
\Big)\,.  \label{selfcons}
\end{eqnarray}
Here, ${\bm v}$ is the Fermi velocity, ${\bm  \Pi} =\nabla +2\pi i{\bm
A}/\phi_0$, ${\bm  A}$ is the vector potential, and $\phi_0$ is the flux quantum. $ \Delta ({\bm r},{\bm k}_F)$ is
 the order parameter that in general depends on the position
${\bm  k}_F$  at the  Fermi surface (or on ${\bm v}$).  The function
$f({\bm  r},{\bm v},\omega) $  originates from Gor'kov's
Green's function  integrated over  energy  near the Fermi surface.
Further,
$N(0)$ is the total density of states at the Fermi level per spin;
the  Matsubara frequencies are $\omega=\pi T(2n+1)/\hbar$
with an integer $n$. The averages over
the Fermi surface are shown as $ \langle... \rangle$.
The   Eilenberger function $g=\sqrt{1-ff^+}=1$ at $H_{c2}$. The temperature $T$ is in energy units, i.e., $k_B=1$.

The self-consistency equation (\ref{selfcons}) is written for
the general case of anisotropic gaps: $\Delta = \Psi({\bf r},T)\,\Omega
({\bf k}_F)$. The function  $ \Omega
({\bf k}_F)$ which determines the $\bm k_F$ dependence of $\Delta$ is normalized so that
 \begin{equation}
\langle\Omega^2\rangle=1\,,
\label{O2=1}
\end{equation}
  for details see, e.g., Ref.\,\onlinecite{RC}. Eq.\,(\ref{selfcons}) corresponds to the factorizable coupling potential, $V(\bm k,\bm k^\prime)=V_0\Omega(\bm k)\Omega(\bm k^\prime)$. This popular approximation\cite{Kad} works well for one band materials with anisotropic coupling. We show in Sections V and VI how this convenient shortcut can be generalized to a multi-band case.

We now recast the self-consistency Eq.\,(\ref{selfcons}) by writing
the solution of Eq.\,(\ref{eil1}) in the form
\begin{equation}
f={2\over \hbar}\int_0^{\infty}d\rho\,e^{-\rho (2\omega +{\bf v}\cdot {\bm
\Pi})} \Delta\,,
\end{equation}
by   using the identity
\begin{equation}
{1\over \hbar \omega}={2\over \hbar}\int_0^{\infty}d\rho\,e^{-2\omega\rho}
\,,
\end{equation}
and by summing up over $\omega$:
\begin{equation}
- \Psi \ln t= \int_0^{\infty} \frac{ du}{\sinh\,u} \Big(
\Psi - \langle\Omega^2 e^{-\rho {\bf v}\cdot {\bm
\Pi}}\Psi\rangle
\Big)\,,  \label{selfcons1}
\end{equation}
where $u=2\pi T\rho /\hbar$ and $t=T/T_c$. Hence, we got rid of the summation over $\omega$, a convenient feature for further analysis.

The self-consistency Eq.\,(\ref{selfcons1}) can be further rewritten in the
form free of the divergent factor $1/\sinh u$ in the integrand.
To this end, we  integrate  by parts
the right-hand side (RHS) of Eq.\,(\ref{selfcons1}) using   $du/\sinh u =d\,\ln \tanh
(u/2)$. The  first term on the RHS  diverges:
\begin{equation}
  \int_0^{\infty} \frac{ du}{\sinh u}  \Psi =-\Psi \ln\tanh\frac{\pi
T\rho}{\hbar}\Big|_{\rho\to 0} \,.  \label{1st}
\end{equation}
The second term transforms to:
\begin{eqnarray}
&-& \Big\langle\Omega^2 \Big[\ln\tanh\frac{\pi
T\rho}{\hbar}\,e^{-\rho {\bf
v}{\bm \Pi}}\Big]_{\rho= 0}^{\infty} \Psi\Big\rangle\nonumber\\
&-& \Big\langle\Omega^2 {\bf v}{\bm \Pi}\int_0^{\infty} d\rho\,
\ln\tanh\frac{\pi T\rho}{\hbar} e^{-\rho {\bf
v}{\bm \Pi}}\Psi\Big\rangle   \,.
\end{eqnarray}
The first term here and the contribution (\ref{1st}) cancel,
and we obtain instead of Eq.\,(\ref{selfcons1}):
\begin{equation}
   \Psi \ln t= \int_0^{\infty}d\rho\, \ln\tanh\frac{\pi T\rho}{\hbar}\,
  \langle\Omega^2 {\bf v}{\bm \Pi}\,e^{-\rho {\bf v}\cdot {\bm
\Pi}}\Psi\rangle
\,.  \label{selfcons3}
\end{equation}
Here, the singularity at $\rho\to 0$ is integrable.

%%%%%%%
\subsection{$\bm{H_{c2}}$ near $\bm{T_c}$}
%%%%%%%

In this domain, the gradients $\Pi\sim \xi^{-1}\to 0$, and one can expand $\exp(-\rho{\bm v}  {\bm \Pi})$ in the integrand (\ref{selfcons3}) and keep
only the linear term:
\begin{equation}
  -\Psi \delta t=  \frac{7\zeta (3)\hbar^2}{16\pi^2T_c^2}\,
   \langle\Omega^2 ({\bm v}\cdot {\bm
\Pi})^2\Psi\rangle \,, \label{GL'}
\end{equation}
where $\int_0^\infty dx\, x\ln \tanh x=-7\zeta(3)/16$ with $\zeta(3)=1.202$.
This is, in fact, the anisotropic version of the linearized Ginzburg-Landau (GL)
equation
\begin{equation}
- \xi^2_{ik}  \Pi_i\Pi_k \Psi =\Psi\,,  \label{GL}
\end{equation}
with
\begin{equation}
  \xi^2_{ik}  = \frac{7\zeta (3)\hbar^2}{16\pi^2T_c^2 \tau}\,
   \langle\Omega^2   v_iv_k \rangle \,,\qquad \tau=1-t\,,     \label{xi}
\end{equation}
the result of Gor'kov and Melik-Barkhudarov.\cite{Gorkov}  Solving the eigenvalue problem for Eq.\,(\ref{GL}) which is similar to one for a charged particle in uniform magnetic field, see e.g. work by  Tilley,\cite{Tilley} one finds
the critical fields in two principal directions of uniaxial materials:
\begin{eqnarray}
& &  H_{c2,c}=\frac{8\pi\phi_0T_c^2\tau}{7\zeta(3)\hbar^2\langle\Omega^2 v_a^2\rangle}\,,\nonumber\\
&& H_{c2,a}=\frac{8\pi\phi_0T_c^2\tau}{7\zeta(3)\hbar^2\sqrt{\langle\Omega^2 v_a^2\rangle\langle\Omega^2 v_c^2\rangle}}    \,,
\label{Hc2-GL}
\end{eqnarray}
so that
 \begin{equation}
\gamma ^2(T_c) =\left(\frac{H_{c2,a}}{H_{c2,c}}\right)^2=
\frac{\xi_{aa}^2}{\xi_{cc}^2}=
\frac{\langle \Omega^2 v_a^2\rangle }{\langle \Omega^2 v_c^2\rangle}\,.
\label{anis_clean}
\end{equation}

The angular dependence
\begin{equation}
H_{c2}(\theta)=\frac{H_{c2,a}}{\sqrt{\sin^2\theta+\gamma^2\cos^2\theta }}\,
\label{Hc2-theta}
\end{equation}
  is a direct consequence of Eq.\,(\ref{GL}) in which $ \xi^2_{ik}$ is a second rank tensor ($\theta$ is the angle between the applied field and the  $c$ axis). We argue below that Eq.\,(\ref{GL}) holds at $H_{c2}$, in fact, at all temperatures, and so should the angular dependence  (\ref{Hc2-theta}), the common practice to call it ``GL" notwithstanding.
  We show that this angular dependence holds for any order parameter symmetry and for any Fermi surface shape  including two-band situations. These conclusions  are, in fact, confirmed experimentally\cite{Terashima,Yuan,Cedomir,CeIrIn5,HHW1,HHW2,Choi1} and by calculations  of Ref.\,\onlinecite{MMK}.

%%%%%%
\section{Isotropic gap  on a Fermi sphere}
%%%%%%

This problem has been solved by HW  for the
whole curve $H_{c2}(T)$.\cite{HW} It is instructive and useful for the following generalization to the anisotropic case to reproduce their result
within the quasiclassic scheme.\cite{coherence}

It was established in Ref.\,\onlinecite{HW}  that at
$H_{c2}(T)$ at any temperature, the order parameter satisfies a linear equation
\begin{equation}
- \xi^2 \Pi^2\Delta=\Delta    \label{e1}
\end{equation}
in which  $\xi(T)$ should be determined so as to satisfy the
self-consistency equation. One can see that this equation is equivalent to the Schroedinger equation for a charge moving in uniform magnetic field and that the maximum field in which non-trivial solutions $\Delta$ exist is $H_{c2}=\phi_0/2\pi\xi^2$.
For the field along $z$ we choose the  gauge $A_y=H x$. One
readily verifies that  in terms of operators
\begin{eqnarray}
\Pi^{\pm}=\Pi_x\pm i\Pi_y,\quad \Pi_x&=&\partial_x,\quad
\Pi_y=\partial_y +iq^2x,\nonumber\\  q^2&=& 2\pi H/\phi_0 \,,
\end{eqnarray}
Eq.\,(\ref{e1})  reads $\Pi^+\Pi^-\Delta=0$ provided
$q^2=1/\xi^2$. Therefore, we obtain a useful property of $\Delta$ at $H_{c2}(T)$:
\begin{equation}
\Pi^-\Delta=\Pi^-\Psi=0\,.\label{property}
\end{equation}
%since $\Omega=1$.

We now introduce $v^{\pm}=v_x\pm iv_y$ so that ${\bm v} {\bm
\Pi}= (v^-\Pi^++v^+\Pi^-)/2$ and evaluate the average $\Big\langle   {\bm v}{\bm \Pi}\,e^{-\rho
{\bm v}{\bm \Pi}}\,\Psi\Big\rangle$ needed in the self-consistency Eq.\,(\ref{selfcons3}). To this end, we use the known property of exponential operators:
\begin{equation}
e^{-\rho {\bm v}{\bm \Pi}}\,\Psi=e^{P+Q}\,\Psi=e^Pe^Qe^{[Q,P]/2}\,\Psi\,.
\label{exp}
\end{equation}
Here, $P=- \rho v^-\Pi^+/2$, $Q=- \rho v^+\Pi^-/2$,
the commutator $ [Q,P] /2=- \rho^2v_{\perp}^2/4\xi^2 $,  and
$v_{\perp}^2= v_x ^2+v_y ^2 $.

Since $\Pi^-\Psi=0$ and $e^Q\Psi=\Psi$, we have:
\begin{eqnarray}
 {\bm v} {\bm \Pi}e^{-\rho {\bm v}{\bm \Pi}}\Psi &=&
 { e^{-\eta}\over 2}(v^-\Pi^++v^+\Pi^-)
\sum_{n=0}^{\infty} \frac{(-\rho v^-\Pi^+)^n}{2^nn!}\Psi ,\nonumber\\
\eta&=&\frac{\rho^2v_{\perp}^2}{4\xi^2}\,. \label{eq20}
\end{eqnarray}
After averaging over the Fermi sphere, only $\langle v^+v^-\rangle$ survives (use $v^\pm =v_\perp e^{\pm i\phi}$):
\begin{equation}
\big\langle    {\bm v}{\bm \Pi}\,e^{-\rho
{\bm v}{\bm \Pi}}\,\Psi\big\rangle =  -{\rho \over
2}\big\langle   e^{-\eta}v^+v^-\big\rangle\Pi^-  \Pi^+\Psi\,.
\label{e37}
\end{equation}
Now, with the help of Eq.\,(\ref{e1}), the self-consistency Eq.\,(\ref{selfcons3})  yields:
\begin{equation}
\xi^2  \ln t= {1\over 2}\int_0^{\infty}d\rho\,\rho\, \ln\tanh\frac{\pi
T\rho}{\hbar}\,
 \left \langle  e^{-\eta}v_{\perp}^2\right\rangle\,.
 \label{eq22}
\end{equation}
Introducing the dimensionless   field
\begin{equation}
h=  \frac{\hbar^2v_F^2}{4\pi^2T_c^2\xi^2}\,.  \label{h}
\end{equation}
and a variable $s=\pi T_c\rho/\hbar$, we rewrite Eq.\,(\ref{eq22}) for $\xi(t)$ as an equation for $h(t)$:
\begin{equation}
  \ln t=   2h\int_0^{\infty}ds\,s\, \ln\tanh(st)
 \left \langle \mu e^{-\mu h s^2}\right\rangle\,,
 \label{selfconsHW}
\end{equation}
where $\mu=v_{\perp}^2/v_F^2$ and $v_F$ is the Fermi velocity. The field $h$ is the upper critical field in units of $\phi_0/ (\hbar^2v_F^2/2\pi T_c^2)$.
Thus, solving Eq.\,(\ref{selfconsHW}) with respect to $h(t)$ we have the HW solution of the problem for the isotropic one-band case. In the following we  refer to Eq.\,(\ref{selfconsHW}) as the HW result although in the original work they obtained a different form of the equation for $H_{c2}(T)$.

At arbitrary $T$, Eq.\,(\ref{selfconsHW}) can be solved  numerically; the exceptions are $T\to 0$ and $T\to T_c$. Since $h(0)$ is finite, the integral over $s$ is truncated at $s\sim 1/\sqrt{\mu h}$; therefore, for low enough $t$ we have $ \ln\tanh(st)\approx \ln t +\ln s$.
 The integration over $s$ then yields:
\begin{equation}
{{\bf C}\over 2}= \Big\langle
\ln \frac{2\pi T_c\xi(0)}{\hbar v_{\perp}}\, \Big\rangle  \,,\label{T=0b}
\end{equation}
where ${\bf C}=0.577$ is the Euler constant.  The averaging over the  Fermi sphere
with $v_{\perp}=v_F\sin\theta$  is readily done:
\begin{eqnarray}
\xi^2(0)=\frac{\hbar^2 v_F^2}{2\pi^2 T_c^2}\,e^{\bm C-2}\,.\label{xi(0)}
\end{eqnarray}
  Thus, we obtain:
\begin{equation}
H_{c2}(0)=\frac{\phi_0}{2\pi\xi^2(0)}=\frac{\phi_0\pi
 T_c^2}{2\hbar^2v_F^2}\,e^{2-{\bf C}}\,,
\end{equation}
the value obtained variationally by Gor'kov\cite{gorkov}
and proven to be exact by HW.

Close to $T_c$, $H_{c2} $ is obtained as   isotropic limits of Eqs.\,(\ref{Hc2-GL}). It is instructive, however, to see how this can be deduced  directly from Eq.\,(\ref{selfconsHW}). In this domain, $\ln t\approx  t-1=-\tau $ and $h \propto\tau$. Then,   the integral in  Eq.\,(\ref{selfconsHW}) should be evaluated in zero order in $\tau$, in other words, $h $ in  $\exp(-\mu h s^2)$ can be set zero. Integration over $s$ gives $-(7\zeta(3)/16) \langle \mu  \rangle$, whereas $\langle \mu  \rangle=\langle v^2_\perp/v_F^2  \rangle=2/3$:
\begin{eqnarray}
h=  12\,\tau/7\zeta(3) \,.   \label{h(Tc)}
\end{eqnarray}
The reduced HW field
\begin{equation}
h^*(0)={H_{c2}(0)\over |H_{c2}^{\prime}(T_c)|T_c}=\frac{h(0)}{|h^\prime(1)|}= \frac
{7\zeta(3) e^2}{48 \, e^{ {\bf C}}} \approx 0.727, \qquad\quad
\end{equation}
  the correct HW value.

%%%%%%
\section{Anisotropic one-band case}
%%%%%%

The central point of the HW paper is the  proof that the linearized GL equation (\ref{e1}) holds not only near $T_c$ but along the whole curve $H_{c2}(T)$. For anisotropic materials Eq.\,(\ref{e1}) should be replaced with Eq.\,(\ref{GL})
where all components of the tensor $(\xi^2)_{ik}$ should be determined from
the self-consistency equation. We consider here
uniaxial materials for which the symmetric tensor $\xi_{ik}^2$ has two
independent eigenvalues.  One has for the field along one of the principal crystal
   directions which we call $z$:
\begin{equation}
-(\xi_{xx}^2 \Pi_x^2 + \xi_{yy}^2\Pi_y^2) \Delta=\Delta\,.
\label{eq28}
\end{equation}
We denote $\xi_{xx} = \xi/\sqrt{m_x}$, $\xi_{yy} = \xi/\sqrt{m_y}$, $\xi_{zz} = \xi/\sqrt{m_z}$ with  dimensionless constants $m_{x,y,z}$.
Since the three quantities, $\xi_{xx}$, $\xi_{yy}$, and $\xi_{zz}$ are replaced
with four, $\xi$  and $m_{x,y,z}$,   we can impose an extra
condition: $m_xm_ym_z=1$. For  uniaxial materials of
interest here with $m_c \ne m_a=m_b$, we introduce the anisotropy
parameter $\gamma = \sqrt{m_c/m_a}$, so that all ``masses" are expressed
in terms of $\gamma$: $m_a=\gamma^{-2/3}$, $m_c=\gamma^{4/3}$. It is worth noting that $m$'s here do not necessarily have the meaning of the band theory effective masses; rather, they are parameters describing the anisotropy of $H_{c2}$; near $T_c$ they can be expressed in terms of Fermi velocities and the gap anisotropy, Eq\,(\ref{anis_clean}).\cite{Gorkov}
Hence, if the ansatz (\ref{GL}) is correct, the self-consistency equation
must provide equations to determine $\xi(T)$ and $\gamma(T)$.

  To make use of the property (\ref{property})  in anisotropic situations
we rescale coordinates in Eq.\,(\ref{eq28}):
\begin{eqnarray}
x'&=&x \sqrt{m_x},\qquad  y'=y\sqrt{m_y}\,,\label{coord}\\
\frac{\partial^2\Delta}{\partial x^{\prime\, 2}}&+&
\Big(\frac{\partial }{\partial y' }
+iq^{\prime\, 2}x'\Big)^2\Delta=-\frac{\Delta}{\xi^2}\,,\label{new GL}\\
q^{\prime\, 2}&=&\frac{2\pi H}{\phi_0\sqrt{m_xm_y}}\,.\label{new-q}
\end{eqnarray}
Formally, Eq.\,(\ref{new GL}) is equivalent to the  isotropic  Eq.\,(\ref{e1}).
The upper critical field is then determined by $q^{\prime\,
2}=1/\xi^2$:
\begin{equation}
H_{c2}=\frac{\phi_0\sqrt{m_xm_y}}{2\pi\xi^2}\,.\label{e28}
\end{equation}
Therefore, we have in the uniaxial case:
\begin{equation}
H_{c2,c}=\frac{\phi_0 }{2\pi\xi^2}\,m_{a }\,,\qquad
H_{c2,b}=\frac{\phi_0}{2\pi\xi^2}\sqrt{m_am_c}\,
\label{eq29}
\end{equation}
and
\begin{equation}
\frac{H_{c2,b}}{H_{c2,c}}=\sqrt{\frac{m_c}{m_a}}=\gamma\,.
\end{equation}
Thus, the formally introduced  ``masses" are related to the measurable ratio of $H_{c2}$'s.

Any coordinate transformation results in transformation of vectors (and tensors).  The scaling transformation (\ref{coord}) necessitates the
   co- and contravariant representations for vectors, see, e.g., Ref.\,\onlinecite{LL2} or \onlinecite{Morse}. The
covariant gradients $\pi_x =\partial/\partial x^{\prime}  $ and $\pi_y
=\partial/\partial y^{\prime} +iq^{\prime\, 2}x'$ have the same
properties as their  isotropic counterparts $\Pi_{x,y}$. Eq.\,(\ref{new GL})  acquires the ``isotropic"  form:
\begin{eqnarray}
&&-\xi^2(\pi_x^2+\pi_y^2)\Delta =\Delta\,,\\
&&\pi^-\pi^+\Delta=-2\Delta/\xi^2\,,\qquad \pi^-\Delta=0\,.\label{e31}
\end{eqnarray}
The following manipulation then is similar to what was done in the isotropic case; one, however, should keep track of differences between co- and contravariant components of vectors.

  Since the scalar products are invariant, we have ${\bm v} {\bm
\Pi}=\nu^i\pi_i=(\nu^-\pi^++\nu^+\pi^-)/2$,   $\nu^\pm=\nu^x\pm i\nu^y$ where \begin{equation}
\nu^x=v_x\sqrt{m_x}\,,\qquad \nu^y=v_y\sqrt{m_y}
\end{equation}
  are the contravariant components of the Fermi
velocity that transform as coordinates, Eq.\,(\ref{coord}).  Further, we use the property (\ref{exp}) of exponential operators
with
\begin{equation}
  P=-{\rho\over 2}\nu^-\pi^+,\, Q=-{\rho\over 2}\nu^+\pi^-,\,
{1\over 2}[Q,P] =-\frac{\rho^2\nu_{\perp}^2}{4\xi^2},
\end{equation}
  and
\begin{equation}
  \quad
\nu_{\perp}^2=(\nu^x)^2+(\nu^y)^2=v_x^2m_x+v_y^2m_y \,.
\end{equation}
Since $\pi^-\Psi=0$ and $e^Q\Psi=\Psi$, we have:
\begin{eqnarray}
&& {\bf v} {\bm \Pi}e^{-\rho {\bf v}{\bm \Pi}}\Psi  \nonumber\\
&&={ e^{-\eta}\over 2}(\nu^-\pi^++\nu^+\pi^-)
\sum_{n=0}^{\infty} \frac{(-\rho\nu^-\pi^+)^n}{2^nn!}\Psi ,\label{eq42}
\end{eqnarray}
where $\eta=\rho^2\nu_{\perp}^2/4\xi^2$.

The next step in the ``isotropic derivation" was to use the fact that $\langle v^+v^-\rangle=2v_F^2/3 $ whereas all other averages (such as $\langle v^+v^-v^-\rangle$) vanish because $ v^\pm=v_\perp e^{\pm i\phi}$ where $\phi$ is the azimuthal angle on a sphere.
To prove rigorously that this is valid for a general Fermi surface is difficult, although it is probably true for uniaxial materials of interest here. As mentioned in the Introduction, to make progress in evaluation of $H_{c2}$ and its anisotropy we   resort to modeling actual Fermi surfaces with spheroids. The rescaling  employed above, in fact, transforms spheroids to ``spheres" in rescaled coordinates so that we can still claim that
 the only surviving product after averaging of Eq.\,(\ref{eq42}) is $\langle\nu^+\nu^-\rangle$ and we obtain:
\begin{equation}
\Big\langle \Omega^2\,  {\bf v}{\bm \Pi}\,e^{-\rho
{\bf v}{\bm \Pi}}\,\Psi\Big\rangle =
  -{\rho \over
2}\big\langle \Omega^2\, e^{-\eta}\nu^+ \nu^-\big\rangle\pi^-  \pi^+
\Psi\,. \label{e37}
\end{equation}
The self-consistency Eq.\,(\ref{selfcons3})  now yields with
the help of Eq.\,(\ref{e31}):
\begin{equation}
\xi^2  \ln t= {1\over 2}\int_0^{\infty}d\rho\,\rho\, \ln\tanh\frac{\pi
T\rho}{\hbar}\,
  \langle\Omega^2  e^{-\eta}\nu_{\perp}^2\rangle
\,.  \label{selfcons4}
\end{equation}
This equation (written for a certain field orientation) contains two
unknown functions, $\xi(T)$ and $\gamma(T)$. Therefore, one has to
write it for two field orientations: (a)  for $\bm H$ along $c$
with $\eta=\rho^2\nu_{\perp}^2/4\xi^2$  and
\begin{equation}
 \nu_{\perp}^2=m_a(v_x^2+v_y^2)=\gamma^{-2/3} (v_x^2+v_y^2)\,,
 \label{nuc}
 \end{equation}
  and (b) for  $\bf H$ along $b$ with
 \begin{equation}
\nu_{\perp}^2= m_av_x^2+m_cv_z^2 =\gamma^{-2/3} (v_x^2+\gamma^2v_z^2)\,.
 \label{nua}
 \end{equation}
  In principle, by solving the system
of these two equations one can determine both $\xi(T)$ and
$\gamma(T)$, thus proving the correctness of the ansatz
(\ref{GL}).

Since the Fermi velocity  is not a constant at anisotropic Fermi surfaces, we  normalize velocities on some value $v_0$ for which we choose
\begin{equation}
v_{0 }^3= \frac{2E_F^2}{\pi^2\hbar^3N(0) } \,,
\label{v0}
\end{equation}
where $E_F$ is the Fermi energy and $N(0)$ is the total density of states at the Fermi level per spin. One easily verifies that $v_0=v_F$ for the isotropic case.

We now write the self-consistency Eq.\,(\ref{selfcons4}) for $\bm H\parallel \bm c$, i.e. with $\nu_\perp$ of Eq.\,(\ref{nuc}), in dimensionless form. To this end, we go to the integration variable $s=\pi T_c/\hbar$, divide both parts by $\xi^2$, and multiply and divide the integrand by $v_0^2$:
\begin{eqnarray}
   \ln t&=& 2 h_c \int_0^{\infty}s\, \ln\tanh (st) \,
 \left \langle\Omega^2 \mu_c  e^{-\mu_c h_c s^2 }\right\rangle ds\,, \qquad
  \label{eq-hc}\\
 h_c&=&\frac{\hbar^2 v_0^2\gamma^{-2/3} }{4\pi^2T_c^2\xi^2}\,, \qquad \mu_c=\frac{v_x^2+v_y^2}{v_0^2} \,.
 \label{mu_c}
\end{eqnarray}
One can  see that $h_c$ is in fact $H_{c2,c}$
in units of $\phi_0/ (\hbar^2v_0^2/2\pi T_c^2)$. An important feature of Eq.\,(\ref{eq-hc}) should be noted: it does not contain the anisotropy $\gamma$ explicitly so that it can be solved for $h_c(t)$.

Writing Eq.\,(\ref{selfcons4}) for $\bm H\perp \bm c$ with   $\nu_\perp$ of Eq.\,(\ref{nua}) we obtain:
\begin{eqnarray}
   \ln t =  2 h_c \int_0^{\infty}s\, \ln\tanh (st) \,
 \left \langle\Omega^2 \mu_b  e^{-\mu_b h_c s^2 }\right\rangle ds\,, \qquad
  \label{eq-gamma}\\
 \mu_b = \frac{v_x^2+\gamma^2v_z^2}{v_0^2} \,.\qquad\qquad\qquad\qquad
 \label{mu_a}
\end{eqnarray}
For the isotropic s-wave  case, $\gamma=\Omega=1$ and Eqs.\,(\ref{eq-hc}) and (\ref{eq-gamma}) coincide with each other and with Eq.(\ref{selfconsHW}) of HW.
   We note that the integrals on the RHS of Eqs.\,(\ref{eq-hc}), (\ref{eq-gamma}) differ only in $\mu$'s;   for brevity we denote:
\begin{eqnarray}
{\cal I}  =\int_0^{\infty}s\, \ln\tanh (st) \,
 \left \langle\Omega^2 \mu   e^{-\mu  h_c s^2 }\right\rangle ds\,.
 \label{cal-I}
 \end{eqnarray}

Thus, the general scheme for solving for $H_{c2}(T)$ and its anisotropy consists of (a) solving Eq.\,(\ref{eq-hc}) for $h_c(t)$  and (b) for the now known $h_c(t)$, solving Eq.\,(\ref{eq-gamma}) for $\gamma(t)$.

 %%%%%%
\subsection{$\bm{T\to T_c}$ and $\bm{T\to 0}$}
%%%%%%

Analysis of Eqs.\,(\ref{eq-hc}), (\ref{eq-gamma}) for arbitrary temperatures is difficult because $h_c(t)$ and $t$ enter the integrals $\cal I$ in a nonlinear fashion. The situation is simpler near $T_c$ where $\ln t\approx -\tau= t-1$ and $h_c\propto\tau$. Therefore, $\cal I$ can be evaluated in zero order in $\tau$, in other words, $h_c$ in the $\exp(-\mu^2_ch_cx^2)$ can be set zero and  $t=1$ in  $ \ln\tanh (xt)$. We obtain after integration:
\begin{eqnarray}
{\cal I} =-\frac{7\zeta(3)}{16} \Big\langle\Omega^2 \mu     \Big\rangle \,, \qquad \label{II-Tc}
\end{eqnarray}
where for $\bm H\parallel \bm c$ one takes  $\mu=\mu_c$ whereas $\mu=\mu_b$ for $\bm H\perp\bm c$. Eq.\,(\ref{eq-hc}) now yields:
\begin{eqnarray}
  h_c= \frac{8\tau}{7\zeta(3)  \left\langle \Omega^2\mu_c \right\rangle } \,,\quad   h_c^\prime(1)= -\frac{8 }{7\zeta(3)  \left\langle \Omega^2\mu_c \right\rangle } \,.\qquad
    \label{hc(Tc)}
 \end{eqnarray}

It is readily shown that Eq.\,(\ref{eq-gamma}) for $\gamma(T_c)$ reduces to
\begin{eqnarray}
  \Big\langle\Omega^2 \mu_b     \Big\rangle = \Big\langle\Omega^2 \mu_c     \Big\rangle\,.
  \label{gam-Tc}
\end{eqnarray}
Using $\mu_b$ and $\mu_c$ of Eqs.\,(\ref{mu_a}), (\ref{mu_c}) one reproduces the general result (\ref{anis_clean}).

As $t\to 0$, $h_c\to$ const, and the exponential factor in $\cal I$ truncates the integrand at a finite $x\sim 1/\sqrt{\mu h_c}$. Hence, at small enough $t$, $\ln\tanh(xt)\approx \ln (xt)=\ln t + \ln x$. One can  now integrate over $x$:
\begin{eqnarray}
  {\cal I} (\mu )= \frac{\ln t }{2h_c}  - \frac{{\bm C}+ \left\langle\Omega^2 \ln (h_c\mu ) \right\rangle  }{4h_c} \,.
    \label{I,t=0}
 \end{eqnarray}
 Substituting this in Eq.\,(\ref{eq-hc}) one obtains:
 \begin{eqnarray}
 h_c(0)=\exp\left(-\bm C -  \big\langle\Omega^2 \ln \mu_c     \big\rangle \right)\,.
  \label{h_c(0)}
\end{eqnarray}
The HW ratio for a general anisotropy reads:
 \begin{eqnarray}
h^*_c(0)= \frac{h_c(0)}{h_c^\prime(1)}=\frac{7\zeta(3)}{8e^{\bm C}}  \left\langle \Omega^2\mu_c \right\rangle \exp \left(- \big\langle\Omega^2 \ln \mu_c     \big\rangle\right) \,.\qquad
  \label{h*}
\end{eqnarray}
Thus, the HW number $h^*(0)=0.727$ is corrected by both Fermi surface shape and by the order parameter symmetry.

  Eq.\,(\ref{eq-gamma}) for $\gamma(0)$ along with expression for $h_c(0)$ readily gives at $T=0$:
\begin{eqnarray}
 \big\langle\Omega^2 \ln \mu_c     \big\rangle =  \big\langle\Omega^2 \ln \mu_b  \big\rangle \,.
  \label{gam(0)}
\end{eqnarray}

%%%%%%
 \section{Two s-wave bands }
 %%%%%%

The general self-consistency equation for two bands reads:
   \begin{equation}
\Delta_\alpha({\bm r},{\bm k})= 2\pi T  \sum_{\beta,\omega} N_\beta \Big\langle V_{\alpha\beta}({\bm k},{\bm k}^\prime) f_\beta ({\bm r},{\bm k}^\prime,\omega) \Big\rangle_{{\bm k}^\prime}\,.
\end{equation}
Here, $\alpha,\beta=1,2$ are band indices and $N_\beta$ are the bands densities of states (DOS').
We consider elements of $V_{\alpha\beta}$ as constants, so that our model is a weak-coupling two-band theory in which the s-wave (i.e., $\bm k$ independent) order parameters $\Delta_{1,2}(\bm r,T,H)$   should be calculated self-consistently for a given coupling matrix $V_{\alpha\beta}$.

As is commonly done, it is convenient to rewrite this equation in the form containing the measured critical temperature which is related to the {\it effective} coupling $V_0$ via the standard BCS formula
 \begin{equation}
 \Delta(0)=\pi e^{-\bm C}T_{c}=2\hbar\omega_D e^{-1/N(0)V_0}\,,
\label{BCS}
\end{equation}
where $\hbar\omega_D$ is the energy scale of a  ``glue boson".
To this end, we introduce the normalized coupling matrix $\lambda_{\alpha\beta}=V_{\alpha\beta}/V_0$ and use the relation identical to Eq.\,(\ref{BCS}) for $T_c$:
\begin{equation}
\frac{ 1}{N(0) V_0}= \ln \frac{T}{T_{c}}+2\pi T\sum_{\omega >0}^{\omega_D}
       \frac{ 1}{\hbar \omega} \,.
\label{1/NV}
\end{equation}
 We then obtain
  \begin{equation}
-\Delta_\alpha \ln t= 2\pi T  \sum_{\omega} \left(\frac{\Delta_\alpha}{\hbar\omega}- \sum_\beta n_\beta\lambda_{\alpha\beta} \Big\langle f_\beta   \Big\rangle\right)\,
\label{s-cons}
\end{equation}
with $n_\beta=N_\beta/N(0)$.

Solving the self-consistency equation in zero field and $\Delta\to 0$ one obtains a relation for $T_c$ (or for $V_0$) in terms of couplings $V_{\alpha\beta}$ and DOS' $n_\alpha$ (Appendix B):
 \begin{eqnarray}
  V_0^{-2} n_1n_2d &-& V_0^{-1} (n_1 V_{11}   + n_2 V_{22})+1=0 , \label{V0}\\
 d& =&  V_{11}  V_{22}-V_{12}^2\,,
 \label{delta}
 \end{eqnarray}
 Therefore, the normalized $\lambda_{\alpha\beta}=V _{\alpha\beta}/V_0$ obey
 \begin{eqnarray}
   n_1n_2 \delta&-&   n_1 \lambda_{11}  - n_2 \lambda_{22} +1=0 , \label{eq94}\\
  \delta& =&  \lambda_{11}  \lambda_{22}-\lambda_{12}^2\,.
 \label{delta}
 \end{eqnarray}
  This property, in fact, means that normalized couplings $\lambda_{\alpha\beta}$ {\it for a given $T_c$} have only two independent components, which are chosen in the following as $\lambda_{11}$ and $\lambda_{22}$.

  To avoid misunderstanding, we stress that   our notation for the normalized  $\lambda_{\alpha\beta}=V _{\alpha\beta}/V_0$ differs from $\lambda_{\alpha\beta} ^{lit}$ used in literature: the latter are $\lambda_{11} ^{lit}=N_1V _{11} $, $\lambda_{22} ^{lit}=N_2V _{22} $, and $\lambda_{12} ^{lit}=N_1V _{12} $.
 It should also be noted that  Eq.\,(\ref{s-cons}) is not valid for the unlikely situation of zero inter-band coupling, $V_{12}=0$, because two decoupled condensates in general have two different critical temperatures.

As was done in Section II, we can transform the self-consistency Eq.\,(\ref{s-cons}) to:
  \begin{equation}
\Delta_\alpha \ln t=\sum_\beta n_\beta\lambda_{\alpha\beta} \int_0^\infty  d\rho\ln\tanh\frac{\pi T\rho}{\hbar}   \Big\langle \bm v\cdot\bm \Pi e^{-\rho\bm v\bm \Pi}\Delta   \Big\rangle_\beta ,\qquad
\label{s-cons2}
\end{equation}
the averaging in the last term here is only over $\beta$-band.

We have generalized the HW isotropic one-band approach by showing that linearized GL equation $-( \xi^2)_{ik} \Pi_i\Pi_k \Delta =\Delta$ holds everywhere along the $H_{c2}$ line. Clearly, the tensor $( \xi^2)_{ik}$ gives the length scales of spatial variations of $ \Delta$ at $H_{c2}$. Before solving  the self-consistency equation for two-band systems (\ref{s-cons2})  it is instructive to recall the situation in the GL domain of a two-band material, which has recently been  discussed in some detail for the case of two isotropic bands.\cite{Zhit,KS,Milosevic}

The system of two GL equations for two order parameters written in terms of coefficients of the GL energy expansion  looks--at first sight--as containing two different coherence lengths, i.e., each order parameter varies in space with its own   length scale different from the other. It has been shown, however, that at $T=T_c$ these two length scales  coincide, provided of course that the system has a single $T_c$.\cite{KS} In fact, two GL equations can be written as one with a single coherence length $\xi$ which is related in a non-trival manner  to coefficients of the GL energy functional. Thus, for a material with two isotropic bands, the linearized GL equation is the same for both bands:
  \begin{equation}
   - \xi^2  \Pi^2 \Delta_\alpha =\Delta_\alpha\,,\qquad\alpha=1,2\,.
 \label{2b-GL}
 \end{equation}

  When the two bands are anisotropic, we can look for solutions of the self-consistency system (\ref{s-cons2}) which satisfies  at $H_{c2}$ the linear equation
\begin{equation}
- (\xi^2)_{ik} \Pi_i\Pi_k \Delta_\alpha=\Delta_\alpha,\qquad  \alpha=1,2 \,. \label{ansatz1}
\end{equation}
All components of the tensor $ (\xi^2)_{ik}$ are to be determined from
the self-consistency equations.  One can consider Eq.\,(\ref{ansatz1}) as an ansatz which should be substituted in the self-consistency relations. If one succeeds in finding such a $ (\xi^2)_{ik}$ so that the latter are satisfied, the ansatz (\ref{ansatz1}) is proven correct.

Repeating the derivation of Section IV we obtain:
  \begin{equation}
   \Big\langle \bm v\cdot\bm \Pi e^{-\rho\bm v\cdot\bm \Pi}\Delta    \Big\rangle_\beta=
\frac{\rho\Delta_\beta}{2\xi^2}\Big\langle \nu_\perp^2 e^{-\eta}  \Big\rangle_\beta  \,  ,\qquad
\label{average}
\end{equation}
where $\xi$ is the average coherence length related to the eigenvalues of $\xi_{ik}^2$: $\xi_{aa}^2=\xi^2\gamma^{2/3}$ and  $\xi_{cc}^2=\xi^2\gamma^{-4/3}$.  Further, $\eta=\rho^2\nu_\perp^2/4\xi^2$ and $ \nu_\perp$ is given in Eq.\,(\ref{nuc}) for $\bm H \parallel  \bm c$ and in
Eq.\,(\ref{nua}) for $\bm H \perp  \bm c$. Substituting this in the system (\ref{s-cons2}) we obtain after straightforward algebra:
\begin{eqnarray}
a_{11}\Delta_1+a_{12}\Delta_2=0\,,\nonumber\\
a_{21}\Delta_1+a_{22}\Delta_2=0\,,
\label{system}
\end{eqnarray}
with
\begin{eqnarray}
a_{11} &=&\xi^2\ln t-\lambda_{11} {\cal J}_1 ,\qquad a_{12}=- \lambda_{12} {\cal J}_2 \,,
 \qquad\nonumber\\
a_{21}&=&- \lambda_{21} {\cal J}_1 \,,\qquad a_{22}=\xi^2\ln t- \lambda_{22} {\cal J}_2 , \qquad
 \qquad \label{aik}\\
 {\cal J}_\alpha&=&\frac{n_\alpha}{2} \int_0^\infty  d\rho\,\rho\ln\tanh\frac{\pi T\rho}{\hbar} \Big\langle \nu_\perp^2 e^{-\eta}  \Big\rangle_\alpha\,. \label{JJ}
\end{eqnarray}
Zero determinant of the  linear system (\ref{system}) gives $\xi(t)$:
\begin{eqnarray}
\xi^4(\ln t)^2&-& \xi^2 \ln t (\lambda_{11} {\cal J}_1+\lambda_{22} {\cal J}_2)
+ \delta{\cal J}_1 {\cal J}_2=0\,.\qquad  \label{det=0}
\end{eqnarray}
For a single band $n_2=0$ and $\lambda_{11}=1$ and  one obtains      Eq.\,(\ref{selfcons4}).

The order parameters $\Delta_{1,2}$ at $H_{c2}$, as solutions of the system (\ref{system}), are determined only within an arbitrary factor, whereas their ratio is fixed: $\Delta_1/\Delta_2=-a_{12}/a_{11}$. This means that $\Delta_1$ and $\Delta_2$ at $H_{c2}$ have the same coordinate dependencies and, in particular, that they have coinciding zeros (this, of course, follows already from Eq.\,(\ref{ansatz1})). The gaps ratio is, in general, temperature dependent (for brevity, we use the term ``gap" instead of ``order parameter" although the latter is more accurate).

Introducing a dimensionless field $h_c$ according to Eq.\,(\ref{mu_c})
one rewrites  Eq.\,(\ref{det=0}) as an equation for $h_c(t)$:
\begin{eqnarray}
 (\ln t)^2&-&  2h_c  (n_1\lambda_{11} {\cal I}_1+n_2\lambda_{22} {\cal I}_2)\ln t \nonumber\\
& +& 4h_c^2(n_1\lambda_{11}+n_2\lambda_{22} -1) {\cal I}_1 {\cal I}_2=0\,, \qquad   \label{det1}\\
{\cal I}_\alpha &=& \int_0^\infty ds\,s\ln \tanh(st)\Big\langle \mu_c  e^{-\mu_c s^2h_c}  \Big\rangle_\alpha\,, \qquad \label{II}
\end{eqnarray}
where $\mu_c$
 is given by Eq.\,(\ref{mu_c}) for the corresponding band,
and we took  account of Eqs.\,(\ref{eq94}) and (\ref{delta}). As in the one-band case, this equation does not contain the anisotropy parameter $\gamma$ and  can be solved numerically for $h_c(t)$.
Equations of a  structure similar to (\ref{det1}) have been employed in studies of  $H_{c2}$   in two-band superconductors.\cite{Gurevich2,Palistrant}

Given $h_c(t)$, one finds the upper critical field:
\begin{eqnarray}
H_{c2,c}=  \frac{2\pi T_c^2\phi_0}{\hbar^2v_0^2}\,h_c(t)\,,\label{H_c2,c}
\end{eqnarray}
where $v_0$ is expressed in terms of the Fermi energy and the total density of states in Eq.\,(\ref{v0}).

 Writing the self-consistency condition for $\bm H\perp \bm c$, we obtain  Eq.\,(\ref{det1}) in which, however, $h_c(t)$ is now known and $\mu_{c,\alpha} $ in  integrals (\ref{II}) is replaced with  $\mu_{b,\alpha}(\gamma)$    given by Eq.\,(\ref{mu_a}) for each band.
    Solving this numerically, one obtains $\gamma(t)$.

The case of  $ T\to T_c $ is treated as was done for a one-band situation:
\begin{eqnarray}
{\cal I}_\alpha \Big |_{T_c} =-\frac{7\zeta(3)}{16} \big\langle \mu  \big\rangle_\alpha  \qquad
 \label{caI_I-Tc_alpha}
\end{eqnarray}
(take Eq.\,(\ref{II-Tc}), set $\Omega=1$ for the s-wave and add  the band index).
For $\bm H\parallel \bm c$ one takes  $\mu=\mu_c$ whereas $\mu=\mu_b$ for $\bm H\perp \bm c$.
The same argument which led to Eq.\,(\ref{I,t=0}) gives for two bands at low temperatures:
\begin{eqnarray}
  {\cal I}_\alpha(\mu )\Big |_{t\to 0}= \frac{\ln t }{2h_c}  - \frac{{\bm C}+ \left\langle \ln (h_c\mu ) \right\rangle_\alpha }{4h_c} \,.
    \label{I_alpha,t=0}
 \end{eqnarray}

One can make progress analytically in looking for $H_{c2}$ near $T_c$ and for $T\to 0$. This calculation is useful for checking the numerical routine; for the sake of brevity we do not provide these somewhat cumbersome results.

%%%%%%%
\subsection{Ratio $\Delta_2 /\Delta_1 $ at $\bm H_{\bm c\bm 2}$}
%%%%%%%

It follows from the system (\ref{system}):
\begin{eqnarray}
\frac{\Delta_1}{\Delta_2} =-\frac{a_{22}}{a_{21}}=\frac{\ln t- 2n_2\lambda_{22}h_c{\cal I}_2}{2n_1\lambda_{12}h_c{\cal I}_1}\,,
 \label{D-ratio}
\end{eqnarray}
where
\begin{eqnarray}
\lambda_{12} =\sqrt{\frac{n_1n_2\lambda_{11}\lambda_{22}-n_1 \lambda_{11}-n_2 \lambda_{22}+1}{n_1n_2}}\,.
 \label{lam12}
\end{eqnarray}
We stress again  that the coordinate independent ratio $\Delta_2(\bm r)/\Delta_1(\bm r)$  makes sense only for order parameters having the same coordinate dependence (in particular, the same zeros and the same  phases).

As $T\to 0$, one can keep only the leading logarithmic term in ${\cal I}_\alpha$ of Eq.\,(\ref{I_alpha,t=0}) to obtain:\cite{remark1}
\begin{eqnarray}
\frac{\Delta_1}{\Delta_2}\Big|_{T=0} = \frac{1-  n_2\lambda_{22} }{ n_2\lambda_{12} }\,.
 \label{D-ratio,T=0}
\end{eqnarray}
We  point out that the zero-$T$  gaps ratio does not depend either on  the Fermi surfaces involved or on the value of $H_{c2}(0)$.  Also, this ratio at $t=0$ is the same for both field orientations;   this is not the case for $t\ne 0$.

 %%%%%%
 \section{Two  bands with gaps of different symmetries}
 %%%%%%

Other than s-wave order parameters emerge if the coupling $V(\bm k,\bm k^\prime)$ responsible for superconductivity is not   a constant (or a 2$\times$2 matrix of $\bm k$ independent constants). The formally simplest way to consider different from s-wave order parameters without going to  details of microscopic interactions is to use a ``separable" potential:
\begin{eqnarray}
 V_{\alpha\beta}(\bm k,\bm k^\prime) = V^{(0)}_{\alpha\beta}\Omega_\alpha(\bm k)\Omega_\beta(\bm k^\prime)\,,
    \label{separable V}
\end{eqnarray}
where $V^{(0)}_{\alpha\beta}$ is a $\bm k$ independent matrix,
and look for the order parameters in the form
\begin{eqnarray}
 \Delta_{\alpha } (\bm r,T,\bm k) = \Psi_{\alpha }(\bm r,T)\Omega_\alpha(\bm k)  \,    \label{separable Delta}
\end{eqnarray}
with the normalization $\big\langle  \Omega ^2    \big\rangle_\alpha=1$ for both bands. One can see that this leads to the self-consistency equation
   \begin{equation}
-\Psi_\alpha\ln t= 2\pi T  \sum_{ \omega}\left(\frac{\Psi_\alpha}{\hbar\omega} -\sum_\beta n_\beta \lambda_{\alpha\beta}\big\langle  \Omega_\beta f_\beta   \big\rangle_\beta\right),
\label{s-cons-new}
\end{equation}
where
\begin{eqnarray}
  \lambda_{\alpha\beta}  = V^{(0)}_{\alpha\beta}/V_0  \,,
  \label{lambda-prime}
\end{eqnarray}
see also Appendix B.  The same algebra as in   Section V results in Eq.\,(\ref{det1}) for $h_c(t)$   with
\begin{eqnarray}
{\cal I}_\alpha = \int_0^\infty ds\,s\ln \tanh(st)\Big\langle \Omega^2\mu_c  e^{-\mu_c h_c s^2}  \Big\rangle_\alpha\,. \qquad \label{II-Omega}
\end{eqnarray}
As before, one calculates the anisotropy $\gamma(t)$ replacing $\mu_c$
with  $\mu_b(\gamma)$.

%%%%%%%%
 \section{Ellipsoid of rotation}
 %%%%%

The Fermi surface as an ellipsoid of rotation is an interesting example on its own right and as a model system for calculating $H_{c2}$ in uniaxial materials with closed Fermi surfaces. Since $H_{c2}$ is weakly sensitive to fine details of Fermi surfaces, calculations done for ellipsoids might be relevant for realistic shapes as well.

Similarly, open Fermi surfaces (extending to boundaries of the Brilouin zone) in uniaxial materials can be  studied qualitatively by considering rotational hyperboloids. The formal treatment of these shapes is similar to that of ellipsoids. This work is still in progress, we show some of it in Appendix D.

Consider an  uniaxial superconductor with the electronic spectrum
\begin{equation}
E(\bm k)=\hbar^2\left(\frac{k_x^2+k_y^2}{2m_{ab}}+\frac{k_z^2}{2m_c}\right)\,,
\end{equation}
so that the Fermi surface is an ellipsoid of rotation with $z$ being the symmetry axis.

In spherical coordinates $(k,\theta,\phi)$ we have
\begin{equation}
E(\bm k)=\frac{\hbar^2k^2}{2m_{ab}}\left(
\sin^2\theta+\frac{m_{ab}}{m_c}\cos^2\theta
\right)=\frac{\hbar^2k^2}{2m_{ab}}\Gamma(\theta)\,,
\end{equation}
so that
\begin{equation}
k_F^2(\theta)=\frac{ 2m_{ab} E_F }{\hbar^2 \Gamma(\theta)} \,.
\label{kF}
\end{equation}
The Fermi velocity is   $\bm v(\bm k)=\bm\nabla_{\bm k}E(\bm k)$,
with the derivatives taken at $\bm k=\bm k_F$:
\begin{eqnarray}
v_x&=&\frac{v_{ab} \sin\theta\cos\phi}{\sqrt{\Gamma(\theta)}}, \,\,\,
v_y=\frac{v_{ab}\sin\theta\sin\phi}{\sqrt{\Gamma(\theta)}},\,\,\,\nonumber \\
 v_z&=&\varepsilon \frac{v_{ab}\cos\theta}{\sqrt{\Gamma(\theta)}},\quad
 \varepsilon=\frac{m_{ab}}{m_c}\,,   \quad v_{ab}= \sqrt{ \frac{2E_F}{m_{ab}} }\,. \quad
\end{eqnarray}
The  value of the Fermi
velocity,  $v =(v_x^2+v_y^2+v_z^2)^{1/2}$, is given by
\begin{equation}
v= v_{ab}\sqrt{\frac{
 \sin^2\theta+\varepsilon^2 \cos^2\theta
}{\sin^2\theta+\varepsilon\cos^2\theta}}=v_{ab}\sqrt{\frac{\Gamma_1(\theta)}{\Gamma(\theta)}} \,.
\end{equation}

The density of states $N(0)$ is defined as an integral over the Fermi surface:
\begin{equation}
N(0)=\int\frac{\hbar^2d^2\bm k_F}{(2\pi\hbar)^3v}  \,.
\label{eqDOS}
\end{equation}
The average  over the Fermi surface can be written as average  over the
solid angle $d\Omega=\sin\theta\;d\theta\, d\phi$:
\begin{equation}
N(0)=
\frac{m^2_{ab} v_{ab}}{2\pi^2\hbar^3}
\int\frac{d\Omega}{4\pi \sqrt{\Gamma(\theta)\Gamma_1(\theta)}}\,.
\label{N(0)}
\end{equation}
The Fermi surface average of a function $A( \theta,\phi)$   is
 \begin{eqnarray}
\langle A( \theta,\phi) \rangle
 =\frac{1}{D}
 \displaystyle\int\frac{d\Omega\,A( \theta,\phi)}{4\pi  \sqrt{\Gamma(\theta)\Gamma_1(\theta)}}\,,\qquad\qquad\label{<A>}\\
D = \int\frac{d\Omega}
{ 4\pi \sqrt{\Gamma(\theta,\varepsilon)\Gamma_1(\theta,\varepsilon)}} =\frac{F(\cos^{-1}\sqrt{\varepsilon},1-\varepsilon)}{\sqrt{1-\varepsilon}}  \qquad
\label{average}
\end{eqnarray}
where $F $ is an Incomplete Elliptic Integral of the first kind. If the function $A$ depends only on the polar angle $\theta$, one can employ the variable $u=\cos \theta$:
\begin{eqnarray}
\langle A( \theta ) \rangle
=\frac{1}{D(\varepsilon)}
 \int_0^1\frac{du\,A(u)}{ \sqrt{\Gamma(u,\varepsilon)\Gamma_1(u,\varepsilon)}}\,,\qquad\qquad\label{<A1>}\\
\Gamma =  1+(\varepsilon-1)u^2\,,\qquad \Gamma_1  = 1+(\varepsilon^2-1)u^2 \,.\qquad
\label{average1}
\end{eqnarray}
It is useful for the following to have a relation between $v_{ab}$ and $v_0$ of  Eq.\,(\ref{v0}) for a one-band situation:
\begin{eqnarray}
v_{ab}^3= D (\varepsilon)\, v_0^3 \,.
 \label{v0-vpar}
\end{eqnarray}

%%%%%%%
\subsection{$\bm H\parallel \bm c$}
%%%%%%%%

  One obtains $\mu_c$ using Eqs.\,(\ref{v0}), (\ref{mu_c}) and (\ref{<A>}):
\begin{eqnarray}
  \mu_ c =D^{2/3}(\varepsilon)\frac{ \sin^2\theta}{\Gamma (\theta,\varepsilon)  }\,.
 \label{mu_c1}
\end{eqnarray}
Hence, we can solve Eq.\,(\ref{eq-hc}) for $h_c(t)$ for a spheroid with a fixed $m_{ab}/m_c=\varepsilon$.
\begin{figure}[htb]
\includegraphics[width=7.cm]{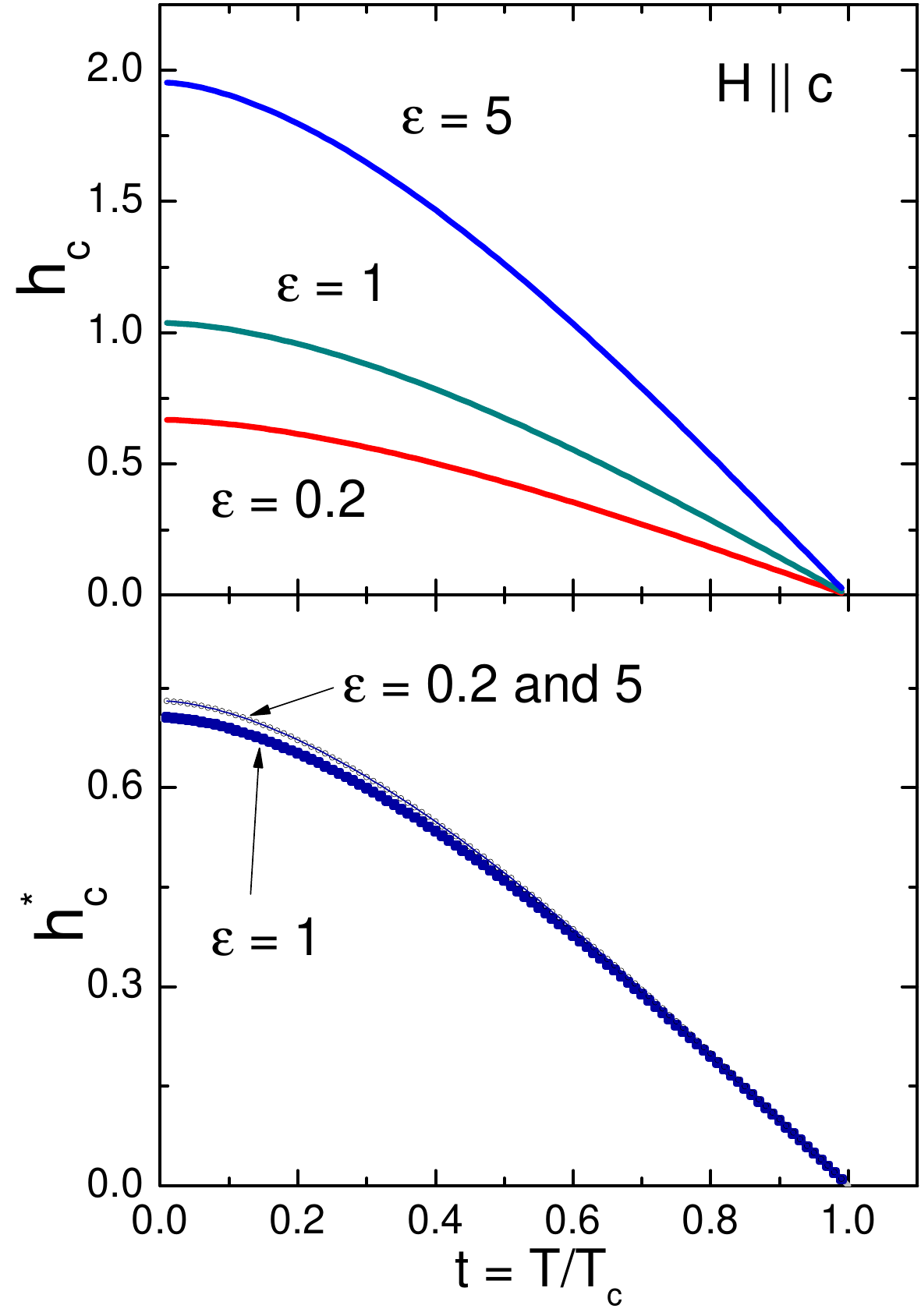}
 \caption{(Color online)  The upper panel: reduced upper critical fields for a prolate $\varepsilon=0.2$ and oblate $\varepsilon=5$ spheroids and s-wave order parameter. $h_c(t)$ is calculated solving Eq.\,(\ref{eq-hc}), (\ref{mu_c}), (\ref{cal-I}). The
HW result for $\varepsilon=1$ is shown for comparison.\\
  The lower panel: the same result plotted using the HW normalization (\ref{h-HW})   to show that in this representation $h_c$ only weakly depends on the Fermi surface shape.
}
 \label{fig1}
 \end{figure}
\begin{figure}[htb]
\includegraphics[width=7.5cm]{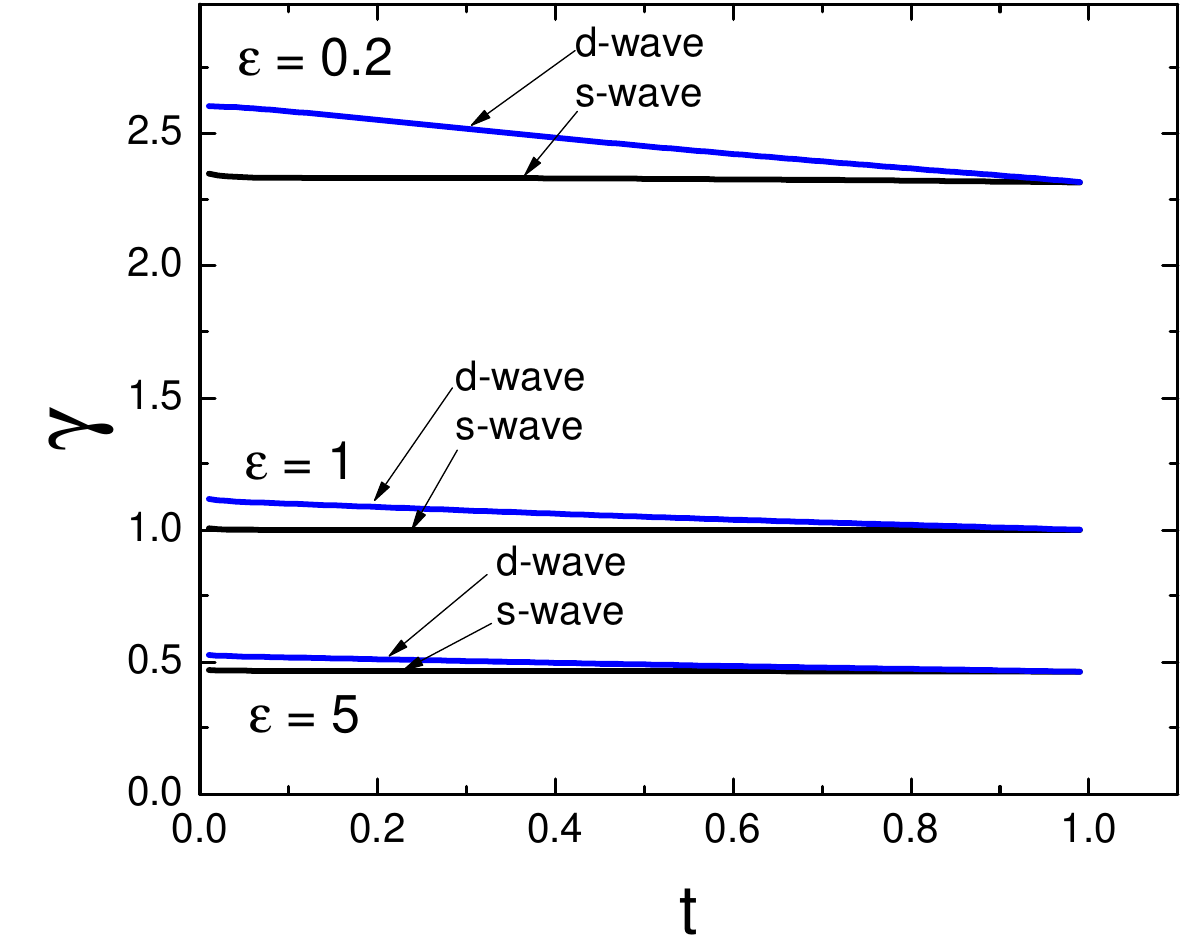}
 \caption{(Color online)   Anisotropy parameter $\gamma=h_{ab}/h_c=H_{c2,ab}/H_{c2,c}$ calculated solving Eq.\,(\ref{eq-gamma}), (\ref{mu_a}), and (\ref{II}) for s- and d-wave order parameters. }
 \label{fig2}
 \end{figure}

Examples of this calculation for the s-wave order parameter, $\Omega=1$,
 are shown in the upper panel of Fig.\,\ref{fig1}  for a prolate ellipsoid with $\varepsilon=0.2$ and for an  oblate one  with $\varepsilon=5$ (the latter corresponds to the ratio of spheroid semi-axes $\sqrt{5}$); the numerical procedure is outlined in Appendix C. The plotted $h_c(t)$ is $H_{c2,c}$ normalized on $\phi_0/ (\hbar^2v_0^2/2\pi T_c^2)$ with $v_0$ given in Eq.\,(\ref{v0}) in terms of the Fermi energy and the total density of states. For comparison, the same results are shown in the lower panel of Fig.\,\ref{fig1}  in the traditional HW normalization
\begin{eqnarray}
  h_c^*= \frac{H_{c2,c}(T)}{T_cH_{c2,c}^\prime(T_c)}= \frac{h_{ c}(t)}{h_{ c}^\prime(1)}\,.
 \label{h-HW}
\end{eqnarray}
It is seen, therefore, that for one-band s-wave materials, although the actual values of $h_c(0)$ vary, the curves of $H_{c2,c}(T)$  have qualitatively similar shapes for different Fermi surfaces.

%%%%%%%%
\subsection{$\bm{\gamma(t)}$}
%%%%%%%%

To solve Eq.\,(\ref{eq-gamma}) for $\bm H\parallel \bm a$, we need
\begin{eqnarray}
  \mu_b  &=&\frac{ v_x^2+\gamma^2v_z^2 }{ v^2_0}\nonumber\\
 & =&D^{2/3}(\varepsilon)\frac{ \sin^2\theta\cos^2\varphi+ \gamma^2 \varepsilon^2\cos^2\theta}{\Gamma (\theta,\varepsilon)}.\qquad
  \label{mu_a1}
\end{eqnarray}

The anisotropy parameter $\gamma$ is calculated with the help of Eq.\,(\ref{eq-gamma}), (\ref{mu_a}) and shown in Fig.\,\ref{fig2}. It is worth observing that for the s-wave case, $\gamma$ depends on the Fermi surface shape but is temperature independent.

One can see that  $\gamma=1$ for a Fermi sphere with $\varepsilon=1$,  as is should be. One can show that   $\gamma( \varepsilon)$   behave approximately as $1/\sqrt{\varepsilon}$. In particular, we observe that for oblate Fermi spheroids, $\gamma < 1$, i.e., $H_{c2,ab}< H_{c2,c}$.

%%%%%%%%
\subsection{d-wave on a one-band ellipsoid}
%%%%%%%%

\begin{figure}[htb]
\includegraphics[width=7.5cm]{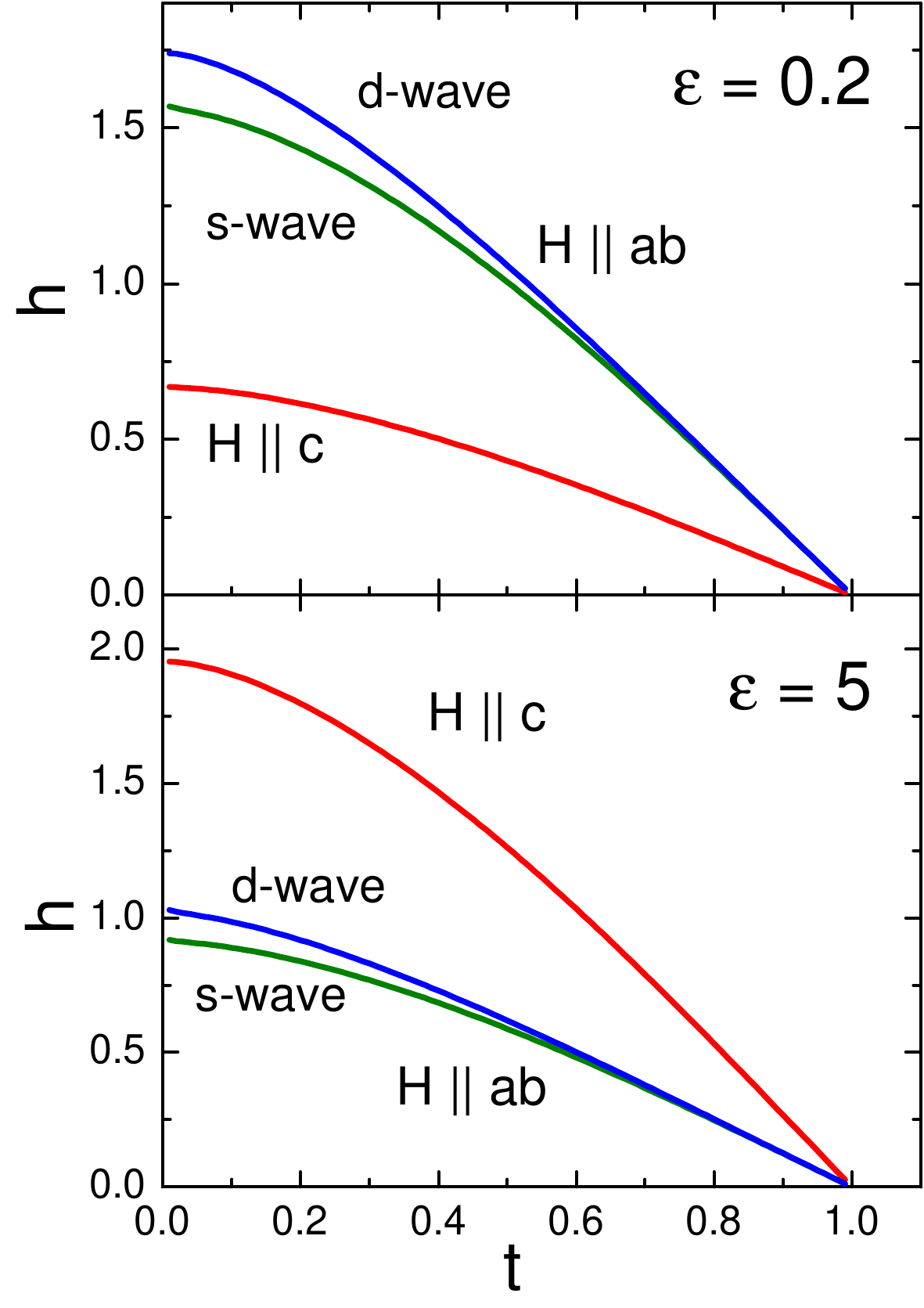}
 \caption{(Color online)   Reduced fields $h(t)$ for two spheroids, $\varepsilon=0.2,5$. For $\bm H\parallel c$, s- and d-curves coincide, as explained in the text. }
 \label{fig3}
 \end{figure}

For this case $\Omega=\Omega_0\cos 2\varphi$ and one can  verify that the condition $\langle\Omega^2\rangle=1$ yields $ \Omega_0^2=2$, the same value  for any Fermi spheroid.  Eq.\,(\ref{eq-hc}) then can be solved numerically with the results shown in Fig.\,\ref{fig3} for values of $\varepsilon$ given in the caption. The anisotropy parameter for d-wave is shown in Fig.\,\ref{fig2}; unlike s-wave, it decreases on warming.

It is worth observing that for a   fixed $\varepsilon$, $h_c(t)$ for  s- and d-wave order parameters are the same. This is because the   Fermi surface average  in Eq.\,(\ref{eq-hc})  involves $(1/2)\int_0^{\pi}\sin\theta d\theta $ for the s-wave whereas for the d-wave we have $(1/4\pi)\int_0^{\pi}\sin\theta d\theta\int_0^{2\pi}2\cos^22\phi\, d\phi$ which has the same value as for the s-case.
 %

%%%%%%%%
\subsection{Order parameter modulated along $\bm{k_z}$}
%%%%%%%%
%
\begin{figure}[htb]
\includegraphics[width=7.5cm]{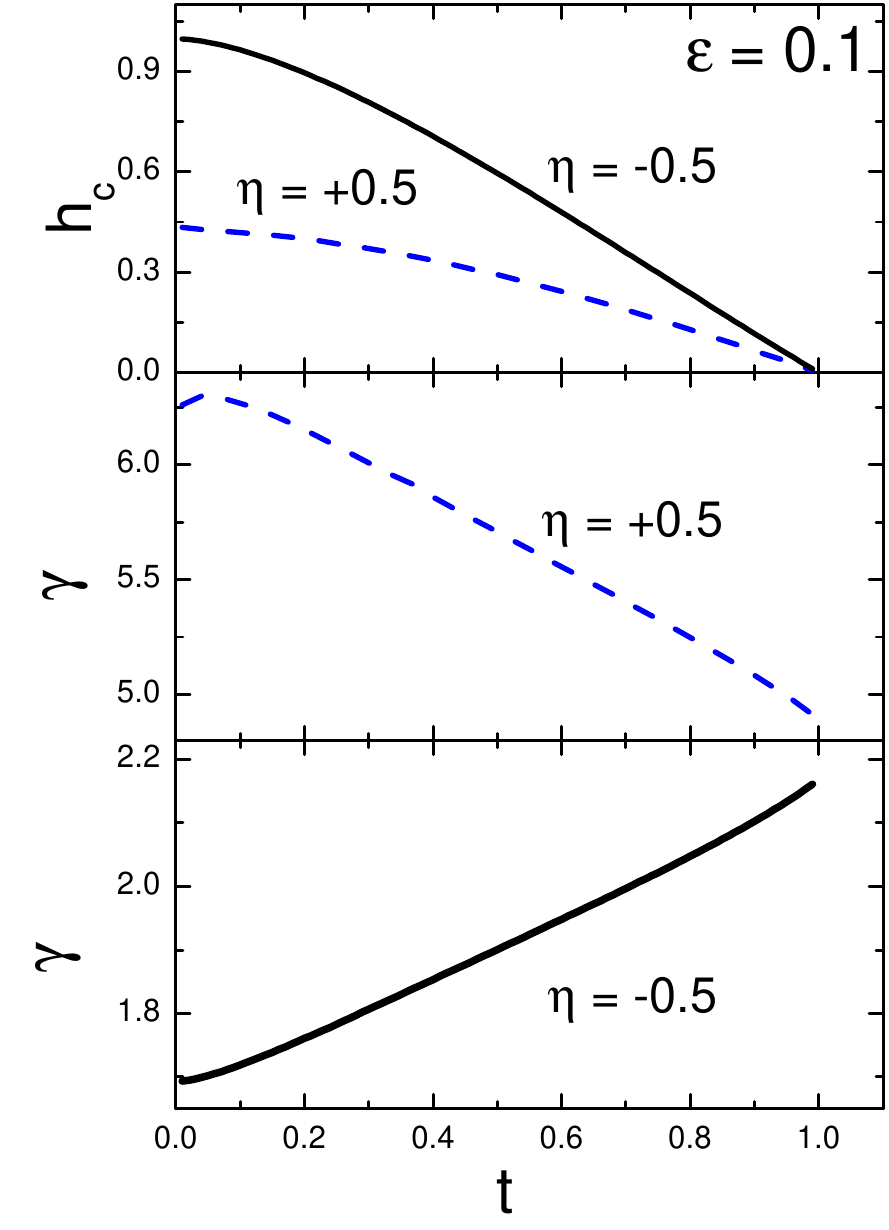}
 \caption{(Color online)  The reduced field  $h_c(t)$ is shown in the upper panel for a spheroid with $\varepsilon=0.1$ and the order parameter (\ref{Xu-form}) with $a=1$  and   $\eta=\pm 0.5$.   Two lower panels show the anisotropy parameter $\gamma(t)$. }
 \label{fig4}
 \end{figure}

The gap function  suggested by the ARPES data for Ba$_{0.6}$K$_{0.4}$Fe$_{2}$As$_{2}$ has a general form:\cite{Xu}
\begin{eqnarray}
 \Delta =  \Delta_0(1+\eta\cos k_za)   \,.
 \label{Xu-form}
\end{eqnarray}
This order parameter varies along the Fermi surface with changing $k_z$; it does not have zeros if $|\eta|<1$. Depending on the sign of $\eta$, it has maximum or minimum at the ``equator" $k_z=0$.

To apply this dependence for  Fermi spheroids, we write:
\begin{eqnarray}
  k_z^2  =k_{F}^2 \cos^2\theta =
   \frac{2m_{ab} E_F \cos^2\theta}{\hbar^2\Gamma(\varepsilon,\theta)}  \,,
 \label{kz}
\end{eqnarray}
where $k_{F}$  is taken from Eq.\,(\ref{kF}). Since $m_\parallel=2E_F/v_\parallel^2$ and $v_\parallel=v_0 D^{1/3}$, we obtain
\begin{eqnarray}
  k_za  =  \frac{2  E_F a\cos \theta}{\hbar v_0D^{1/3}(\varepsilon)\sqrt{\Gamma(\varepsilon,\theta)}}  \,.
 \label{kza}
\end{eqnarray}
We now choose the length scale
\begin{eqnarray}
   a  =  \frac{\hbar v_0 }{2  E_F  }  \,,
 \label{a}
\end{eqnarray}
so that
\begin{eqnarray}
  k_za  =  \frac{ \cos \theta}{ D^{1/3}(\varepsilon)\sqrt{\Gamma(\varepsilon,\theta)}}  \,.
 \label{kza1}
\end{eqnarray}
Note that in the isotropic case, the length $a=a_0/(3\pi^2)^{1/3}$ with $a_0$ being the interparticle spacing (the unit cell size).

To adopt the order parameter (\ref{Xu-form}) for our formalism we define
\begin{eqnarray}
  \Omega^2   =  \frac{(1+\eta\cos (k_za))^2}{ \langle(1+\eta\cos (k_za))^2\rangle}  \,,
 \label{Om^2}
\end{eqnarray}
as to satisfy the normalization $ \langle\Omega^2\rangle=1$
(the average in the denominator is calculated according to Eq.\,(\ref{<A1>})).

Numerical evaluation of $h_c(t)$ for parameters given in the caption of Fig.\,\ref{fig4} result in a curve qualitatively similar to that of HW (the upper panel), however, the anisotropy parameter  decreases on warming for $\eta >0$ as shown in the middle panel. It is quite remarkable that changing the sign of $\eta$, i.e., going from order parameters with a maximum at the Fermi surface ``equator"  $k_z=0$ ($\eta >0$) to ones with a minimum ($\eta <0$), not only changes the temperature dependence of $\gamma$ to the opposite, but affects its absolute values   as well (for positive $\eta$, the anisotropy parameter is noticeably larger than for $\eta<0$ for other parameters kept the  same).

One can readily evaluate how the anisotropy of the London penetration depth $\Lambda$,  $\gamma_\Lambda=\Lambda_c/\Lambda_{ab}$, changes with temperature  for   $\eta=-0.5$ (for which the $H_{c2}$ anisotropy is shown in the lowest panel of Fig.\,\ref{fig4}). To this end we note that  the GL theory requires the same values of these two anisotropies at $T_c$, so that  $\gamma_\Lambda(T_c)=\gamma (T_c)\approx 2.17$ according to Fig.\,\ref{fig4}. As $T\to 0$, $\gamma_\Lambda^2(0)\to\langle v_x^2\rangle / \langle v_z^2\rangle$ (in the clean limit, the order parameter does not enter the anisotropy of the London depth).\cite{RC,gamma-model,RPP} The calculation of these averages is straightforward for a spheroid with $\varepsilon=0.1$: $\gamma_\Lambda(0)\approx 3.38$. Thus, the $\Lambda$-anisotropy decreases on warming, unlike $H_{c2}$-anisotropy. This qualitative behavior of the $\Lambda$-anisotropy is, in fact, seen in experiments on Ba(Fe$_{1-x}$Co$_x$)$_2$As$_2$, (Ba$_{1-x}$K$_x$)Fe$_2$As$_2$, and NdFeAs(O$_{1-x}$F$_x$).\cite{Prozetal}

\begin{figure}[htb]
\includegraphics[width=7.5cm]{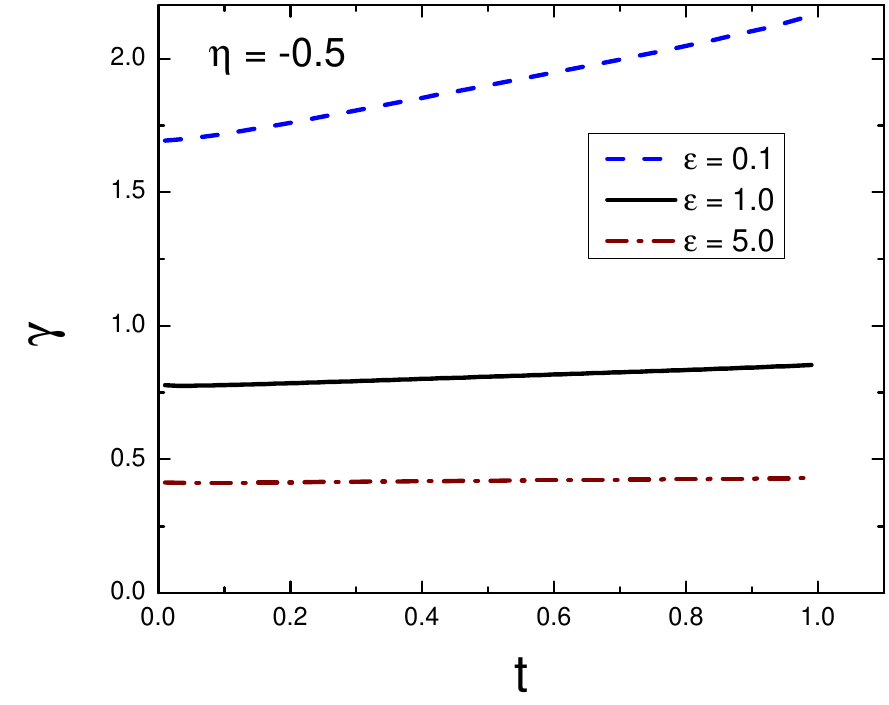}
 \caption{(Color online)      The anisotropy parameter $\gamma(t)$ for three different Fermi surfaces $\varepsilon=0.1,1,5$ and the   order parameter of the form (\ref{Xu-form}) with $\eta=-0.5$. }
 \label{fig4a}
 \end{figure}

In Fig.\,\ref{fig4a} we demonstrate that the effect of $\gamma$ increasing on warming remains if the Fermi surface changes from a prolate spheroid with $\varepsilon=0.1 $ to a sphere $\varepsilon=1 $ and to oblate spheroid with $\varepsilon=5$.
In particular, these features challenge  the common belief that temperature dependence of the anisotropy parameter is always related to a multi-band situations.

We note in concluding this Section that other possible anisotropic order parameters can be treated within our scheme in a similar manner.

%%%%%%%%%
\section{Two-band results}
%%%%%%%%%

To apply the theory developed  for two-band materials one first should map actual band structure upon two spheroids, the procedure we demonstrate in some detail on the well-studied  MgB$_2$. When calculating parameters $\mu_{c }$ (and $\mu_b$) needed in this mapping for each  band, one should bear in mind that   in the two-band situation we have:
\begin{eqnarray}
  \mu_ {c,\alpha} =\left[\frac{D(\varepsilon_\alpha)}{n_\alpha}\right]^{2/3} \,\frac{ \sin^2\theta}{\Gamma (\theta,\varepsilon_\alpha)  }\,,
 \label{mu_calpha}
\end{eqnarray}
see Appendix C.

%%%%%%%%
\subsection{MgB$_2$}
%%%%%%%%

We take this example to demonstrate  that our procedure yields $H_{c2,c}(T)$ and the anisotropy $\gamma(T)$ in agreement with existing data (see, e.g.  Refs.\,\onlinecite{Lyard,Budko}) and with calculations of Refs.\,\onlinecite{MMK,DS,Palistrant}.
We stress that our calculations of $H_{c2}$   are done with the same set of coupling parameters as those used for  the zero-field properties of this material as described in Ref.\,\onlinecite{gamma-model}.
\begin{figure}[htb]
\includegraphics[width=7.5cm]{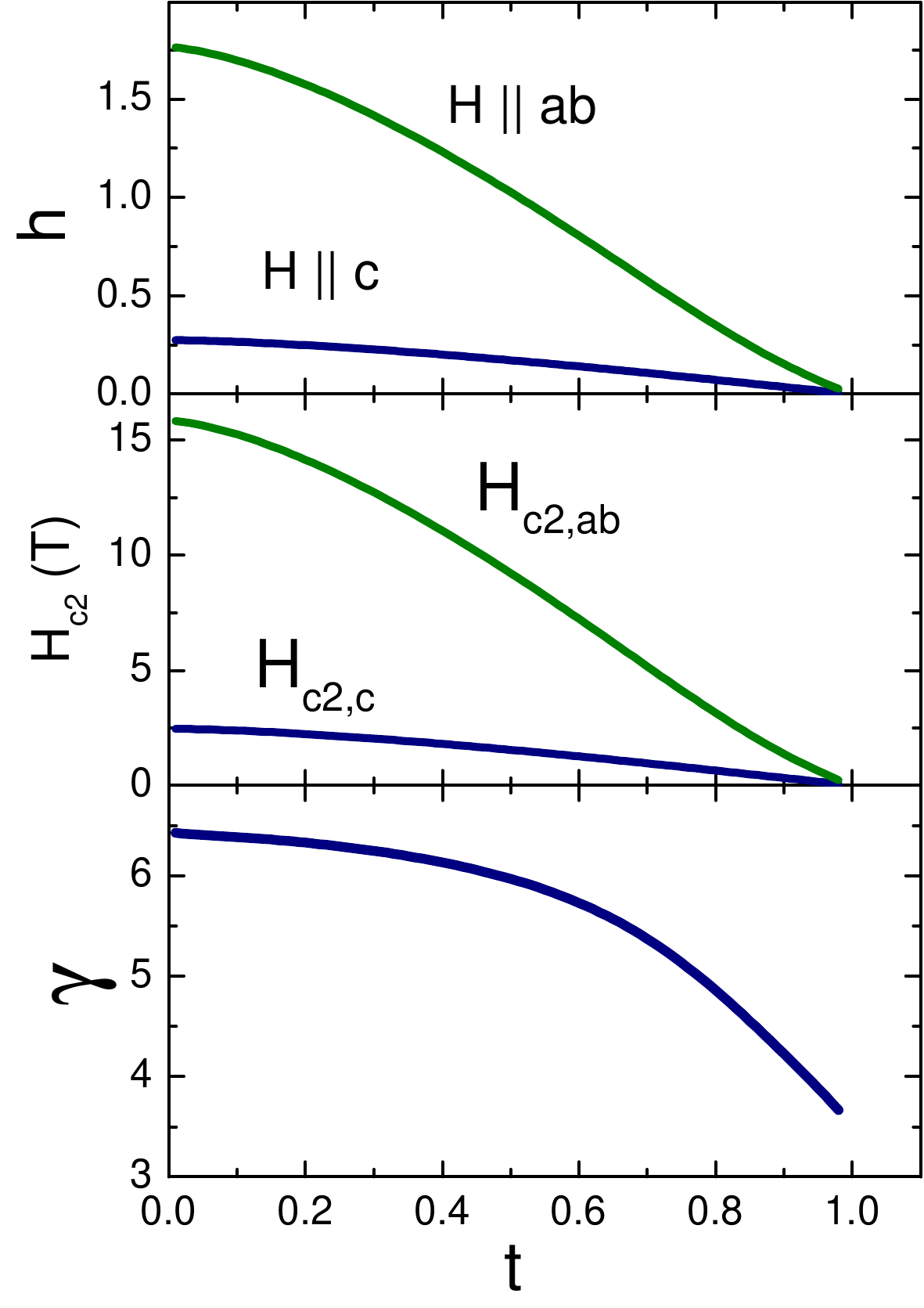}
  \caption{(Color online)  The {\it upper} panel: reduced upper critical fields of clean MgB$_2$. $h_c(t)$ is calculated solving Eq.\,(\ref{det1}), (\ref{II}) for two spheroids with ellipticity parameters $\varepsilon_1=0.02867$ (a strongly prolate spheroid for nearly 2D $\sigma$ band) and $\varepsilon_2=1.273$ (for a  3D $\pi$ band) evaluated with the help of two-bands Fermi velocities from the band structure calculations of Ref.\,\onlinecite{Bel}. The reduced couplings $\lambda_{\alpha\beta} $ are evaluated on the basis of microscopic calculations of Ref.\,\onlinecite{Golubov} as described in the text.
 \indent  The {\it middle} panel: the same for the fields in common units;  $H_{c2,c}(0)\approx 2.8\,$T and $H_{c2,ab}(0)\approx 16\,$T.
  \indent   The {\it lower} panel: the anisotropy parameter $\gamma=h_{ab}/h_c=H_{c2,ab}/H_{c2,c}$ calculated solving Eq.\,(\ref{det1}), (\ref{II}) with $\mu_c$ replaced by $\mu_b(\gamma)$. }
 \label{fig5}
 \end{figure}
 The four Fermi sheets of MgB$_2$  can be grouped in two effective bands with  nearly constant zero-field gaps for each group.\cite{Choi} The two effective bands are mapped here upon two ellipsoids.\cite{MMK} We describe this   procedure   in some detail.

We take the following data from the band structure calculations:\cite{Golubov}
the relative densities of states of our model are
$n_1\approx 0.42$ and $n_2\approx 0.58$    for $\sigma$ and $\pi$ bands, respectively.\cite{Bel,Choi}
The band  calculations\cite{Bel} provide also the averages over
separate Fermi sheets: $\langle v_a^2\rangle_1=23$,
 $\langle v_c^2\rangle_1=0.5$, and
$\langle v_a^2\rangle_2=33.2$, $\langle v_c^2\rangle_2=42.2
\times 10^{14}\,$cm$^2$/s$^2$.

To map this system onto two Fermi  ellipsoids, we note that the averages over  spheroids are given by
\begin{eqnarray}
 \langle v_x^2\rangle   &=& \frac{v_\parallel^2}{2D(\varepsilon)} \int_0^1\frac{du(1-u^2)}{\Gamma ^{3/2}(\varepsilon)\Gamma_ 1
 ^{1/2} (\varepsilon)}
 \,,\label{<vx2>}\\
 \langle v_z^2\rangle   &=& \frac{v_\parallel^2\varepsilon^2}{D(\varepsilon)} \int_0^1\frac{du \,u^2 }{\Gamma ^{3/2}(\varepsilon)\Gamma_{1 }^{1/2}(\varepsilon)} \label{<vx2>}\,,\quad\qquad
 \label{<vz2>}
  \end{eqnarray}
  where $u=\cos\theta$. The integrals here can be expressed in terms of Elliptic Integrals, alternatively they can be evaluated  numerically.
 Forming the ratio $ \langle v_x^2\rangle/ \langle v_z^2\rangle$ we obtain an equation which can be solved  for $\varepsilon$. This gives  $\varepsilon_1=0.02867$  and $\varepsilon_2=1.273$. For a given $\varepsilon $ and, e.g.,  $\langle v_z^2\rangle\equiv\langle v_c^2\rangle$, we obtain $v_{ab}^2$ for  two ellipsoids: $v_{ab,1}^2=0.6019\times 10^{14}$ and $v_{ab,2}^2=1.436\times 10^{16}$\,(cm/s)$^2$. Next, we write:
\begin{eqnarray}
  \frac{1}{v_0^3}&=&\frac{\pi^2\hbar^3(N_1+N_2)}{2E_F^2}=  \frac{1}{v_{01}^3}+ \frac{1}{v_{02}^3}\nonumber\\
   &=&  \frac{D(\varepsilon_1)}{v_{\parallel,1}^3} + \frac{D(\varepsilon_2)}{v_{\parallel,2}^3}
   \label{1/v03}
  \end{eqnarray}
where Eq.\,(\ref{v0-vpar}) has been used; this gives $v_0^2=3.867\times 10^{14}$\,(cm/s)$^2$, the constant used in the field normalization.

To obtain normalized coupling constants
  $\lambda_{\alpha\beta}=V_{\alpha\beta}/V_0$ we use the effective  values (calculated including Coulomb repulsion)   which in our notation read: $N_1V_{11}=0.807$, $N_2V_{22}=0.276$, $N_1V_{21}=0.118$, $N_2V_{12}=0.086$.\cite{Golubov} Using Eq.\,(\ref{V0}) one evaluates $1/V_0N(0)= 1.211$, and obtains the normalized coupling
  \begin{eqnarray}
  \lambda_{11}= \frac{V_{11}}{V_0}=\frac{N_1V_{11}}{N(0)V_0}\frac{1}{n_1}\approx 2.328\,.  \label{lam11}
  \end{eqnarray}
Similarly, we find $\lambda_{22}=0.5765$, $\lambda_{12}=  \lambda_{21}= 0.340$.
With these input parameters we have solved Eqs.\,(\ref{det1}),  (\ref{II}) for $h_c(t)$.    The result is shown in the upper panel  of Fig.\,\ref{fig5}.

Given $h_c(t)$, we rewrite the same equations where $\mu_c$ is replaced with $\mu_b(\gamma)$ and solve them for $\gamma(t)$. The latter is shown in the lower panel  of Fig.\,\ref{fig5}.

Hence, the calculation gives $h_c(0)\approx 0.292$. We now use the relation (\ref{H_c2,c}) between the dimensionless $h_c$ and physical $H_{c2,c}$, where $T_c=39.5\,$K and $v_0$ has been estimated above, to obtain $H_{c2,c}(0)\approx 2.8\,$T, the value close to the observed.\cite{Budko,Lyard} The general behavior of $H_{c2,c}(T)$ is close to that of HW, the fact confirmed by experiment. It should be mentioned that our result is close to the calculations of Miranovic et al,\cite{MMK}  Dahm and Schopohl,\cite{DS} and Palistrant.\cite{Palistrant}  We, however, do not reproduce the calculated $H_{c2,c}(T)$ with changing curvature and  a substantial upturn at low $T$'s of Ref.\,\onlinecite{Gurevich2}.
 \begin{figure}[htb]
 \includegraphics[width=7.5cm]{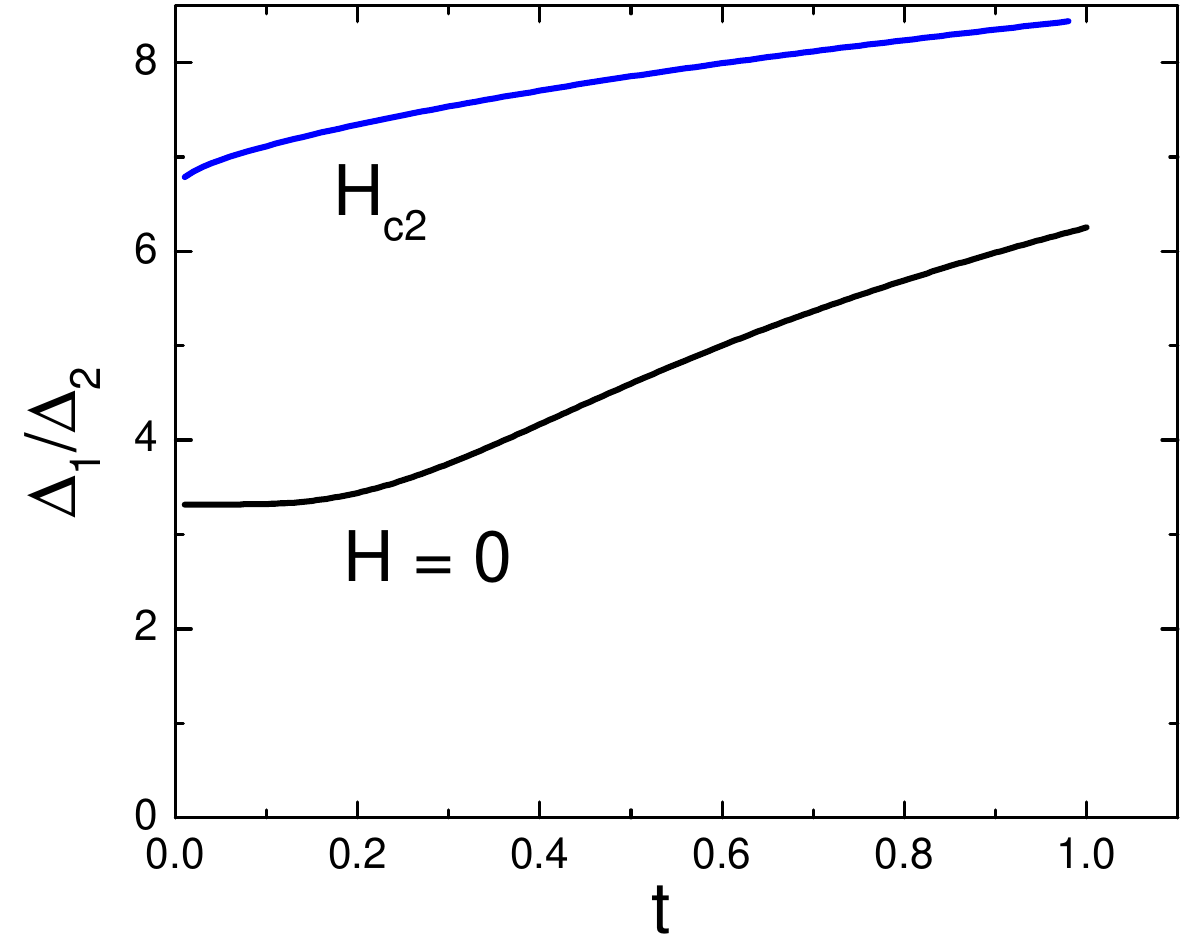}
 \caption{(Color online)   The ratio $\Delta_1/\Delta_2$ of two order parameters of MgB$_2$ at  $H_{c2,c}$ and at zero field  calculated with the same coupling constants.
 }
 \label{fig6}
 \end{figure}

 The general formula for the gaps ratio at $H_{c2}$ is given in Eq.\,(\ref{D-ratio}). Figure \ref{fig6} shows this ratio  as a function of temperature at $H_{c2,c}$. For comparison we  show the gaps ratio in zero field calculated with the help of the same coupling constants.\cite{gamma-model} It is instructive to note that the gaps ratio at $H_{c2}$  exceeds the zero-field value at all temperatures, that can be interpreted as a stronger suppression of the small gap by the magnetic field than that of the large one at the leading $\sigma$ band. %We also note the absence of a low-$T$ plateau of the zero-field ratio in the ratio at  $H_{c2}$.
  At  first sight, one would expect the two ratios to coincide as $T\to T_c$, which our results clearly do not show. Such an expectation, however, would   not be justified: even when  $H_{c2}\to 0$, the superconductor in the mixed state differs from the uniform state by an extra magnetic field suppression of the order parameter.

One often finds  in literature   the statement  that a small gap in MgB$_2$ is substantially or even completely suppressed by a large enough field. This would correspond to a substantial increase of $\Delta_1/\Delta_2$ or even divergence of this ratio at some field under $H_{c2}$.
Our result,  however, shows that even at $H_{c2}$ the gaps ratio is finite and of the same order at all $T$'s. We conclude that the full suppression of the small gap does not happen at any field   $H \le H_{c2}$.  On the other hand, assuming (as was done in Ref.\,\onlinecite{MMK}) that the gap ratio at $H_{c2}$ is the same as that calculated in zero field, is also incorrect.

 %%%%%%%%
\subsection{$\bm{\lambda_{11}\sim\lambda_{22}\ll | \lambda_{12}|}$ }
%%%%%%%

This case  is close to theoretical models of pnictides in which the interband coupling is assumed dominant. We consider here a limiting situation $ \lambda_{11}=\lambda_{22}=0$ to simplify the algebra. Indeed, for the two field directions we have:
\begin{eqnarray}
 (\ln t)^2&-&   4h_c^2 \, {\cal I}_1(\mu_c )\, {\cal I}_2(\mu_c )=0\,, \qquad   \label{eq-c}\\
(\ln t)^2&-&   4h_c^2 \, {\cal I}_1(\mu_b)\, {\cal I}_2(\mu_b)=0\,. \qquad   \label{eq-a}
\end{eqnarray}
The first equation here can be solved for $h_c(t)$. Since $\mu_b$ depends on $\gamma$,   the second gives an equation for $\gamma(t)$. The latter can also be written in a different form if one subtracts Eq.\,(\ref{eq-c}) from  (\ref{eq-a}):
\begin{eqnarray}
  {\cal I}_1(\mu_c ) \,{\cal I}_2(\mu_c )=  {\cal I}_1(\mu_b)\, {\cal I}_2(\mu_b) \,.     \label{difference}
\end{eqnarray}
  \begin{figure}[h ]
  \includegraphics[width=7cm]{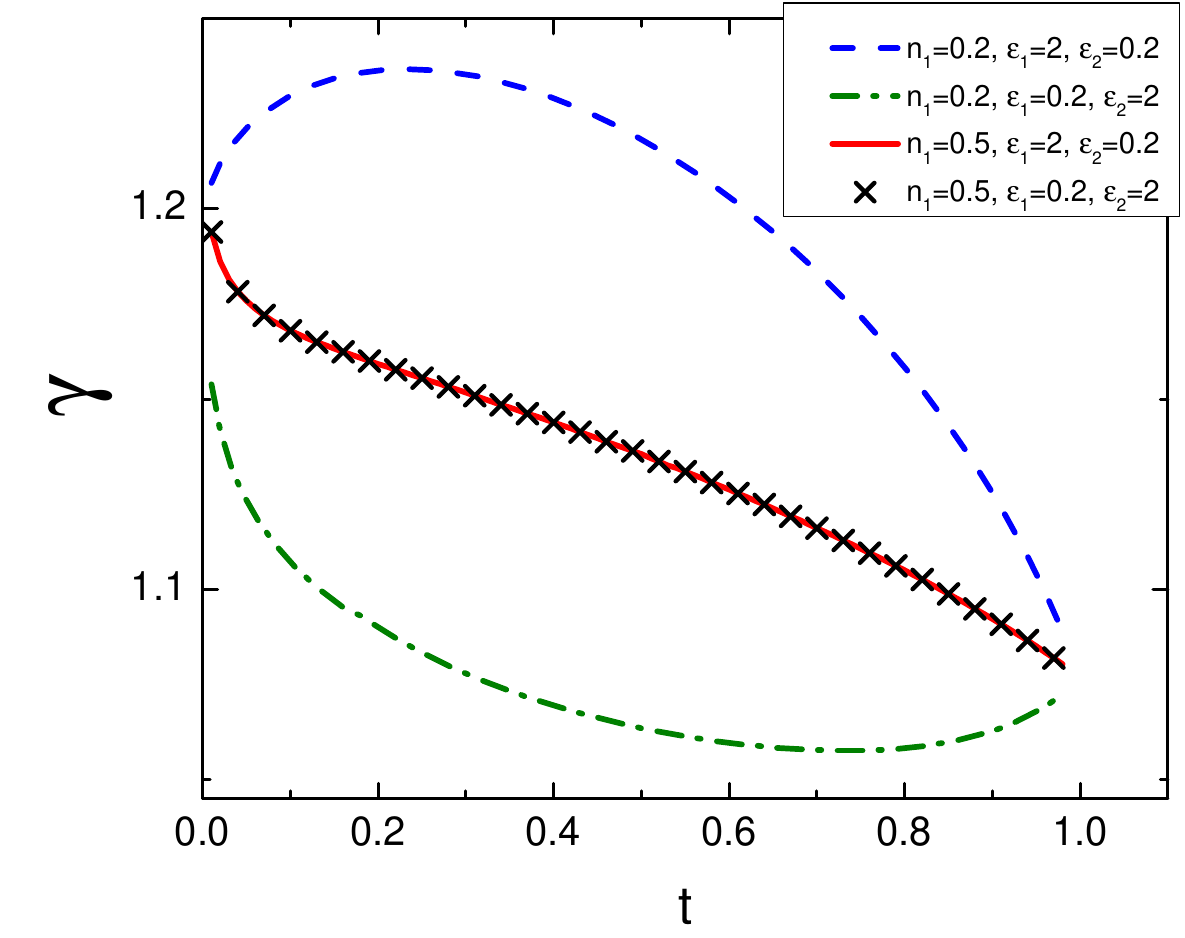}
 \caption{(Color online)  The field $h_c(t)$ (the upper panel) and the anisotropy parameter $\gamma(t)$ (the lower panel) for the order parameter of the form $ \Delta =  \Delta_0(1+\eta\cos k_za)$ . The parameters chosen are: $\varepsilon_1=\varepsilon_2=0.1$,   $n_1=0.5$, $a=1$, for three sets of $\eta$'s given in the figure.}
 \label{fig7}
 \end{figure}
Fig.\,\ref{fig7} shows  numerical solution for  $\gamma(t)$  for a few representative parameter sets. It is worth noting the particularly informative feature of Fig.\,\ref{fig7}: it shows that the anisotropy $\gamma(t)$ is not necessarily a monotonic  function of temperature.
 We note that numerical calculations of $\gamma(t)$ showing an extremum at $0<t<1$ are quite robust.

%%%%%%%
\subsubsection{$\bm{\Delta_1/\Delta_2}$}
%%%%%%%%

The general result for the gaps ratio (\ref{D-ratio}) gives for $\lambda_{11}= \lambda_{22}=0$:
\begin{eqnarray}
\frac{\Delta_1}{\Delta_2} =\frac{\ln t }{2n_1\lambda_{12}h_c{\cal I}_1}\,.
 \label{D-ratio1}
\end{eqnarray}

Near $T_c$ we use $\cal I$'s of Eq.\,(\ref{II-Tc}) and $h_c$ from Eq.\,(\ref{eq-c}) to obtain:
\begin{eqnarray}
\frac{\Delta_1}{\Delta_2}\Big|_{T_c} = \frac{1}{ n_1\lambda_{12} }\sqrt{ \frac{ \langle \Omega^2\mu_c   \rangle_2}{ \langle \Omega^2\mu_c     \rangle_1} } = \sqrt{\frac{n_2 \langle \Omega^2\mu_c   \rangle_2  } {n_1 \langle \Omega^2\mu_c  \rangle_1} }\,,
 \label{D-ratio2}
\end{eqnarray}
since for $\lambda_{11}=\lambda_{22}=0$, the normalized $\lambda_{12}^2=1/n_1n_2$.
As $T\to 0$, we can keep only the leading term $\sim \ln t$ in $\cal I$ of Eq.\,(\ref{I,t=0}):
\begin{eqnarray}
\frac{\Delta_1}{\Delta_2}\Big|_{T=0} =\sqrt{\frac{ n_2}{n_1} }\,.
 \label{D-ratio3}
\end{eqnarray}

 It is  instructive to note that the last relation in the form
$n_1\Delta_1^2= n_2\Delta_2^2$ at $T=0$ suggests the equipartition of condensation energy between the bands (provided the only non-zero coupling is $\lambda_{12}$).

%%%%%%%
\subsubsection{Two bands with $\bm{\Delta( k_z)}$}
%%%%%%%

\begin{figure}[h ]
  \includegraphics[width=7cm]{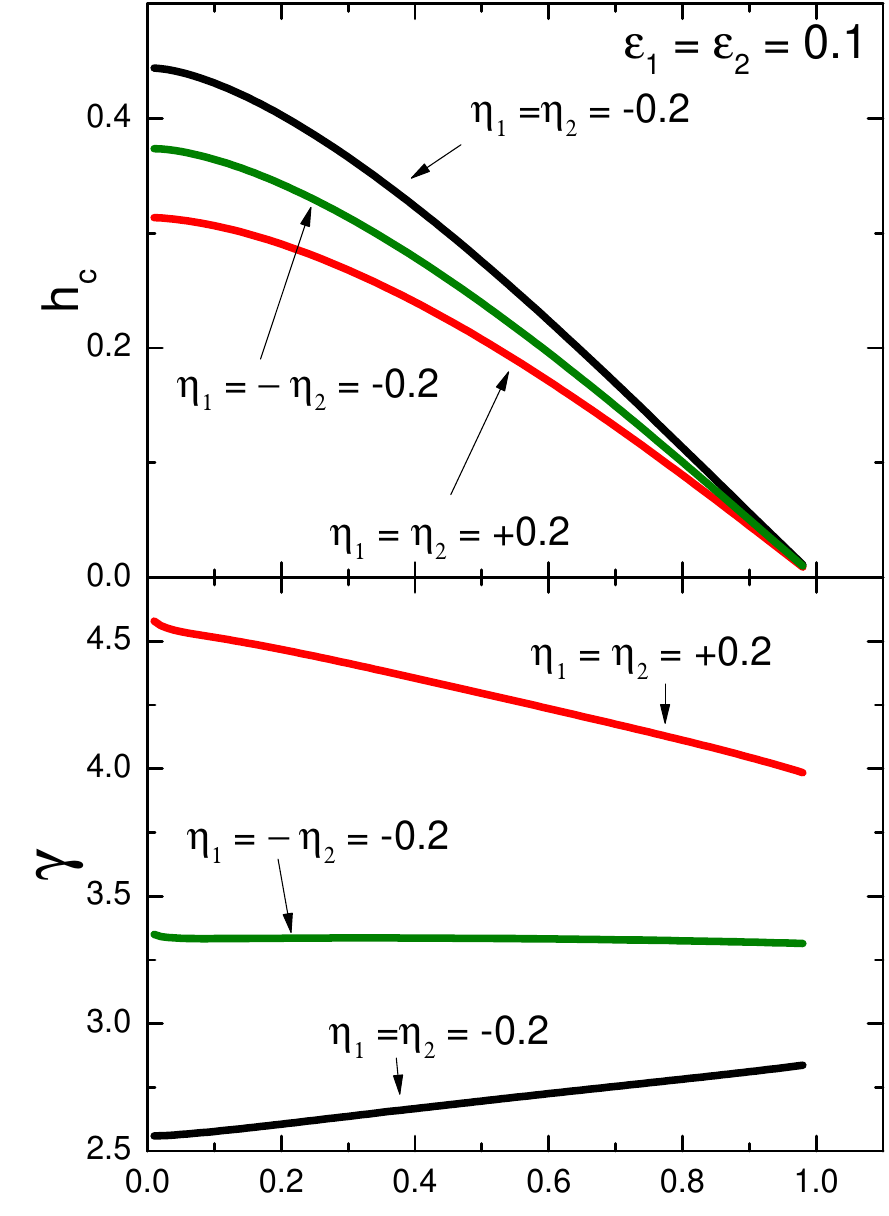}
 \caption{(Color online)  The field $h_c(t)$ (the upper panel) and the anisotropy parameter $\gamma(t)$ (the lower panel) for the order parameter of the form $ \Delta =  \Delta_0(1+\eta\cos k_za)$ . The parameters chosen are: $\varepsilon_1=\varepsilon_2=0.1$,   $n_1=0.5$, $a=1$, for three sets of $\eta$'s given in the figure.}
 \label{fig8}
 \end{figure}

This example is of interest because it may have implications for understanding the behavior of $H_{c2}(T)$ and, in particular, its anisotropy in Fe-based materials. In Fig.\,\ref{fig8} we show $h_c(t)$ and $\gamma(t)$ for two nearly cylindrical bands with order parameters modulated along $k_z$ according to Eq.\,(\ref{Xu-form}).  Modulations are characterized by $\eta_1=\eta_2=  -0.2$; other parameters are given in the caption. The main feature worth paying attention to is the  anisotropy $\gamma(t)$ which is increasing on warming for  negative $\eta$'s.

It is worth noticing that experimental anisotropy of Fe-based materials behaves qualitatively similar to that shown by the lowest curve in the lower panel of Fig.\,\ref{fig8}.\cite{Budko-Canfield} We, however, do not have enough information to fix the necessary parameters for  realistic calculations (one needs partial densities of states,  Fermi surfaces characterized separately for relevant bands for evaluating the geometric parameters $\varepsilon$'s, and the order parameters $\Delta(\bm k_F)$). Hence, we take our results as having a generic qualitative value. In particular, among various possibilities we have considered, only the order parameters of the form $ \Delta =  \Delta_0(1+\eta\cos k_za)$ combined with dominant role of the inter-band coupling  yield the $\gamma(t)$ increasing on warming similar to what is seen experimentally.

%%%%%%%%
 \section{Summary and Conclusions}
%%%%%%%%

  The upper critical field $H_{c2}$ and its anisotropy are among  the easiest properties to examin when a new superconductor is discovered. This work provides a relatively straightforward scheme for evaluating the orbital $H_{c2}(T)$ and its anisotropy  $\gamma(T)$ for single and two band uniaxial materials. We reproduce here the main points of our approach.

   The input parameters  for two band materials are (i) the coupling matrix $V_{\alpha\beta}$ (or the normalized couplings $\lambda_{\alpha\beta}$ and $T_c$), (ii) the symmetries of the order parameter on two bands   given as $\Omega_\beta(\theta,\phi)$, $\beta$ is the band index, with the normalization (\ref{O2=1}), and (iii) the characteristics of electron systems (Fermi surfaces, averages of squared Fermi velocities, DOS').

The equation to solve for the reduced upper critical field $h_c(t)$ parallel to the $c$ crystal axis of a two-band clean uniaxial material reads:
\begin{eqnarray}
 (\ln t)^2&-&  2h_c  (n_1\lambda_{11} {\cal I}_1+n_2\lambda_{22} {\cal I}_2)\ln t \nonumber\\
& +& 4h_c^2(n_1\lambda_{11}+n_2\lambda_{22} -1) {\cal I}_1 {\cal I}_2=0\,, \qquad   \label{119}\\
{\cal I}_\beta &=& \int_0^\infty ds\,s\ln \tanh(st)\Big\langle \Omega^2\mu_c  e^{-\mu_c s^2h_c}  \Big\rangle_\beta \,, \qquad \label{120}
\end{eqnarray}
where   $\mu_c$
 is given by Eq.\,(\ref{mu_c}) for the corresponding band, and $\Omega$'s describe  the order parameter symmetries. After $h_c(t)$ is found, one solves Eq.\,(\ref{119}), in which $\mu_c$   is replaced by $\mu_b(\gamma)$ of Eq.\,(\ref{mu_a}), for  $\gamma(t)$.
  The one-band case is obtained by setting $n_2=0, n_1=1$ and $\lambda_{11}=1$. The case of two s-wave bands corresponds to $\Omega_1=\Omega_2=1$.

Because $H_{c2}$ is determined by equations containing only  integrals over the Fermi surface, it is insensitive to fine details of the Fermi surface shape. Therefore, one can replace actual Fermi surfaces with ellipsoids (or with spheroids, for uniaxial materials).
  Given the averages of the squared Fermi velocities over each band, one establishes  the geometry of corresponding rotational ellipsoids  (the squared ratio  of semi-axes, $\varepsilon$). This procedure is described in Section VIII.A on the well-studied example of MgB$_2$.

  One numerically solves Eqs.\,(\ref{119}),  (\ref{120})
 by employing any of available packages (such as Mathematica)   able to find  roots of nonlinear equations and to do multiple integrals.
 The scheme can also be applied  to the case of two bands with order parameters of  different symmetries.

By design, our method is applicable for clean materials with a moderate $H_{c2}(0)$; paramagnetic limiting effects are out of the scope of this work.\cite{Gurevich2}
The method differs from those previously employed by not involving  explicit  coordinate dependent $\Delta_{1,2} $ and minimization relative to the vortex lattice structures  in calculating $H_{c2}(T)$.\cite{Scharn-Klemm,Rieck,MMK,DS} The main feature of the two-band derivation is that the linearized GL equation (\ref{ansatz1}) is assumed to hold at $H_{c2}(T)$ at any $T$, the ansatz proven correct by satisfying the self-consistency equation of the theory.
The method is tested on the well-studied example of MgB$_2$ where it shows a satisfactory agreement with data and with other calculations.

Our main results are as follows:

1. We find that in  clean one-band s-wave materials, the $T$ dependence of the anisotropy $\gamma$ cannot be caused by the Fermi surface anisotropy (however, the paramagnetic limit, which is out of the scope of this paper, may suppress $\gamma$ at low $T$'s and cause $\gamma(T)$ to increase on warming).

2. For other than s-wave symmetry,  $\gamma $  depends on temperature even for one-band materials. This dependence  is pronounced for open Fermi surfaces as well as  for order parameters depending on $k_z$. Thus,  the common belief that the temperature dependence of the anisotropy parameter is always related to multi-band situations is incorrect.

3. Our scheme of calculating $H_{c2}$ for two-band materials does not utilize any assumptions about the field and temperature dependences  of the order parameters $\Delta_\alpha$ in two bands.\cite{MMK} In fact, the gaps ratio is calculated self-consistenly and in general turns out temperature dependent.
Although both $\Delta_ 1$ and $\Delta_ 2$ turn zero at $H_{c2}(T)$, their ratio is finite and in the examples examined is larger than at zero field.

 4. The case  of exclusively  inter-band  coupling  is discussed, that might be relevant while interpreting   data on $H_{c2}$ and its anisotropy in Fe-based compounds.

5. For order parameters of the form $ \Delta =  \Delta_0(1+\eta\cos k_za) $ (one of the candidates suggested for pnictides),  the anisotropy parameter $\gamma(t)$ depends on the sign of $\eta$ (or $\eta$'s for two-bands).  In particular, $\gamma(T)$ increases on warming in a nearly linear fashion (as  for pnictides) for both $\eta$'s negative.

\section*{Acknowledgements}

Some ideas described in this text were conceived in discussions with Predrag Miranovich while working on  $H_{c2}(T)$ of MgB$_2$ in 2002; VK is grateful  to Predrag for this experience. We thank Andrey Chubukov for turning our attention to order parameters of the form (\ref{Xu-form}) and  our Ames Lab colleagues John Clem, Andreas Kreyssig, Sergey Bud'ko, Makariy Tanatar and Paul Canfield for interest and critique. We are grateful  to Erick Blomberg for reading the manuscript and useful remarks.
Work at the Ames Laboratory is supported by the Department of Energy - Basic Energy Sciences under Contract No. DE-AC02-07CH11358.

%%%%%%%
\appendix
%%%%%%%

 %%%%%%
 \section{Different form of the one-band equation for ${\bm H_{\bm c\bm 2}}$}
 %%%%%%%
Both sides of Eq.\,(\ref{eq-hc})  diverge logarithmically when $t\to 0$, so that these divergences, in fact, cancel out. However, in numerical work this cancellation may not always be exact which may cause  the numerical solutions to become unreliable in this limit. Here, we provide an alternative form of this equation free  of this shortcoming.

To this end, consider an identity:
\begin{eqnarray}
 \ln  t&=&\left \langle \Omega^2 2 h_c \mu  \int_0^\infty ds\, s \ln (st) \,
   e^{-\mu  h_c s^2 }\right\rangle \nonumber\\
  & +&\frac{\bm C+\left \langle\ln(\Omega^2h_c\mu)   \right\rangle}{2}  \,,
  \label{eqB1}
\end{eqnarray}
which is verified   by direct integration. We now combine it with Eq.\,(\ref{eq-hc}):
\begin{eqnarray}
&&\frac{\bm C+\left \langle\Omega^2\ln(h_c\mu)   \right\rangle}{4  h_c } \nonumber\\ &&=     \int_0^\infty ds\, s \ln \frac{ \tanh(st)}{st} \left \langle \Omega^2\mu
   e^{-\mu  h_c s^2 }\right\rangle.\qquad
  \label{eqB2}
\end{eqnarray}
As $t\to 0$, the integral on the RHS goes to zero, and we immediately obtain
the result (\ref{h_c(0)}) for $h_c(0)$.

%%%%%%%%%
\section{${\bm T_{\bm c}}$ as a function of ${\bm\lambda_{\bm \alpha\bm \beta}}$}
%%%%%%%%%

This question had been discussed,   e.g., in Ref.\,\onlinecite{gamma-model}). Since our notation of normalized $\lambda_{\alpha\beta}$ differs from that in the literature, we provide here corresponding relations. The s-wave self consistency equation for $H=0$,
\begin{eqnarray}
 \Delta_ \nu  =2\pi T\sum_{\mu,\omega}  N_\mu V_{\nu\mu} f_\mu(\omega)\,,\qquad
  \label{B1}
\end{eqnarray}
gives near $T_c$ where $ f_\mu = \Delta_ \mu/\hbar\omega$:
\begin{eqnarray}
 \Delta_ \nu  =\sum_{\mu }   n_\mu N(0)V_0\lambda_{\nu\mu}  \Delta_  \mu\sum_\omega^{\omega_D}\frac{2\pi T}{\hbar\omega}\,,\qquad
  \label{B2}
\end{eqnarray}
where $V_0$ is   to be defined. We choose it so that
\begin{eqnarray}
 \frac{1}{ N(0)V_0}=\sum_\omega^{\omega_D}\frac{2\pi T}{\hbar\omega}=\ln\frac{2\hbar\omega_D}{1.76 T_c}\,,\qquad
  \label{B3}
\end{eqnarray}
or, which is the same, $1.76 T_c=2\hbar\omega_De^{-1/N(0)V_0}$. We then obtain a linear and homogeneous system of equations $ \Delta_ \nu  =\sum_{\mu }   n_\mu  \lambda_{\nu\mu}  \Delta_  \mu$, zero determinant of which gives Eq.\,(\ref{eq94}).

For other than s-wave order parameters on two bands, we take the coupling potential in the form (\ref{separable V}) and the order parameters as in
Eq.\,(\ref{separable Delta}), We then obtain the self-consistency equation $ \Psi_ \nu  =2\pi T\sum_{\mu,\omega}  N_\mu V^{(0)}_{\nu\mu} \langle \Omega_\mu f_\mu(\omega)\rangle$. We now denote
$  \lambda_{\nu\mu}  = V^{(0)}_{\nu\mu}/V_0 $, Eq.\,(\ref{lambda-prime}),
and recall that $f_\mu=\Omega_\mu \Psi_\mu/\hbar\omega$ near $T_c$. This gives $ \Psi_ \nu  =\sum_{\mu }   n_\mu  \lambda_{\nu\mu}  \Psi_ \mu$, i.e. the same system of equations as above for the s-wave case and the same zero-determinant condition (\ref{eq94}) albeit with renormalized couplings (\ref{lambda-prime}).

\section{$\bm \mu_{\bm c}$ for the two-band case}

By definition,  $\mu_{c,\beta}\propto v^2_{ab,\beta}/v_0^2$  with $v_0$ given in Eq.\,(\ref{v0}). For $v_{ab,\beta}$ we have two relations: one with the Fermi energy, $m_{ab,\beta}v_{ab,\beta}^2=2E_F$, and the other in terms of the band DOS, Eq.\,(\ref{N(0)}):
 \begin{equation}
N_\beta=\frac{m^2_{ab,\beta} v_{ab,\beta}}{2\pi^2\hbar^3}
\,D(\varepsilon_\beta)\,,\qquad \beta=1,2\,.
\label{N_alpha}
\end{equation}
One excludes $m_{ab,\beta}$ from these two relations to obtain:
\begin{equation}
v_{ab,\beta}^3= \frac{2E_F^2}{\pi^2\hbar^3N_\beta } \,.
\label{v||}
\end{equation}
Hence, we have:
\begin{equation}
\mu_{c,\beta}=\frac{v_{ab,\beta}^2}{v_0^2}= \left[\frac{D(\varepsilon_\beta) }{n_\beta}\right]^{2/3} \,,
\label{v-ratio}
\end{equation}
a clear generalization of Eq.\,(\ref{v0-vpar}) for the one-band spheroid. This gives Eq.\,(\ref{mu_calpha}) used in the text.

 %%%%%%
 \section{Open Fermi Surface}
 %%%%%%%

\begin{figure}[h]
   \includegraphics[width=5cm]{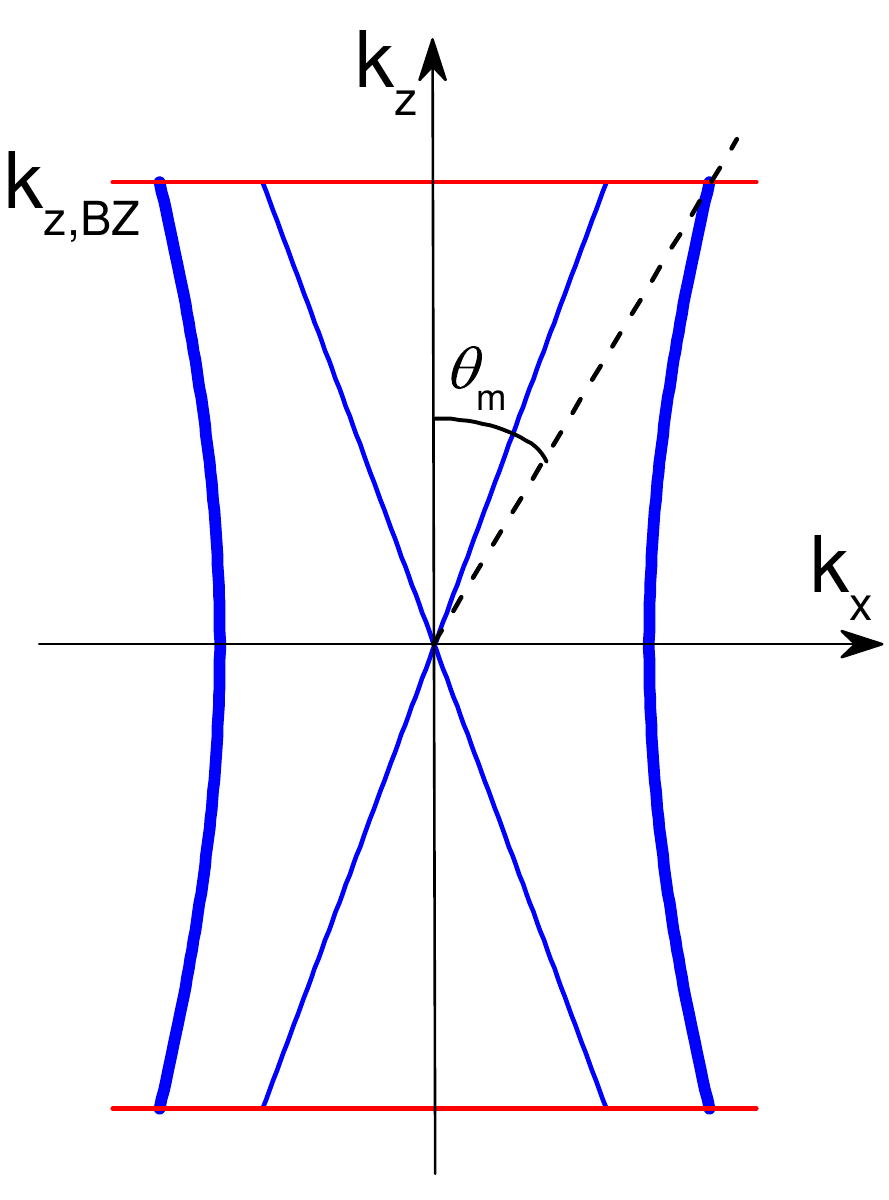}
 \caption{(Color online)  The cross-section $k_y=0$ of an open  Fermi hyperboloid. $k_{z,BZ} $ is   the zone boundary in the $\hat c$ direction. Tilted straight lines correspond to $E=0$, whereas the hyperbola shows the Fermi surface. }
 \label{hyperb}
 \end{figure}

 The theory employed above is  designed  to model Fermi surfaces closed within the first Brillouin zone. Here we consider an example of the Fermi surface which crosses the zone boundary, i.e. it is open. Perhaps, the simplest shape to consider is a rotational hyperboloid which is a property of the carriers energy of the form:
 \begin{eqnarray}
E(\bm k)=\hbar^2\left(\dfrac{k_x^2+k_y^2}{2m_{ab}}-\dfrac{k_z^2}{2m_c}\right)\,.
\label{spectrum}
\end{eqnarray}
A schematic picture is shown in Fig.\ref{hyperb}.
 In spherical coordinates $(k,\theta,\phi)$ we have
\begin{eqnarray}
E(\bm k)&=&\dfrac{\hbar^2k^2}{2m_{ab}}\left(
\sin^2\theta-\dfrac{m_{ab}}{m_c}\cos^2\theta
\right)=\dfrac{\hbar^2k^2}{2m_{ab}}\Gamma_2(\theta),\qquad\nonumber\\
\Gamma_2 &=& \sin^2\theta-\epsilon\cos^2\theta\,,\quad \epsilon = m_{ab}/m_c\,.
\end{eqnarray}
It is seen from the figure that in the first quadrant of the plane $k_x,k_z$, the Fermi surface is situated at $\theta>\theta_m>\tan^{-1}\sqrt{\epsilon}$, i.e., everywhere at the Fermi surface $\Gamma_2>0$.

The angle $\theta_m$ corresponds to the crossing of the Fermi hyperboloid with the zone boundary $k_{z,BZ}=2\pi/c$ where $c$ is the unit cell size along the $\hat c$ direction:  \begin{eqnarray}
 \tan^2\theta_m =\epsilon+\frac{m_{ab} c^2E_F}{2\pi^2\hbar^2} =\epsilon +\alpha\,.
 \label{theta_m}
\end{eqnarray}
% It is easy to see that   $\alpha= \tan^2\theta_{cyl}$ for a cylindrical Fermi surface ($m_c \to\infty$, $\epsilon\to 0$).
The parameter  $ \alpha = k ^2_F(\pi/2)/k_{z,BZ}^2\,$  in most situations of interest is less than unity ($k_{F}(\pi/2)$ is the radius of the hyperboloid neck).

The Fermi momentum is given by
\begin{equation}
k_F^2(\theta)=\frac{ 2m_{ab} E_F }{\hbar^2 \Gamma_2(\theta)} \,.
\label{kF}
\end{equation}
The Fermi velocity is   $\bm v(\bm k)=\bm\nabla_{\bm k}E(\bm k)_{\bm k_F}$:
%with the derivatives taken at $\bm k=\bm k_F$:
\begin{eqnarray}
v_x=\dfrac{v_{ab} \sin\theta\cos\phi}{\sqrt{\Gamma_2(\theta)}}, \,\,\,
v_y&=&\dfrac{v_{ab}\sin\theta\sin\phi}{\sqrt{\Gamma_2(\theta)}},\nonumber \\
 v_z=\epsilon \dfrac{v_{ab}\cos\theta}{\sqrt{\Gamma_2(\theta)}},\qquad
  v_{ab}&=& \sqrt{ 2E_F/m_\parallel} \,.\qquad
\end{eqnarray}
  Futher, we have for
  $v =(v_x^2+v_y^2+v_z^2)^{1/2}$:
\begin{equation}
v= v_{ab}\sqrt{\dfrac{
 \sin^2\theta+\epsilon^2 \cos^2\theta
}{\sin^2\theta-\epsilon\cos^2\theta}}=v_{ab}\sqrt{\dfrac{\Gamma_1(\theta)}{\Gamma_2(\theta)}} \,.
\end{equation}
  \begin{figure}[t]
   \includegraphics[width=7cm]{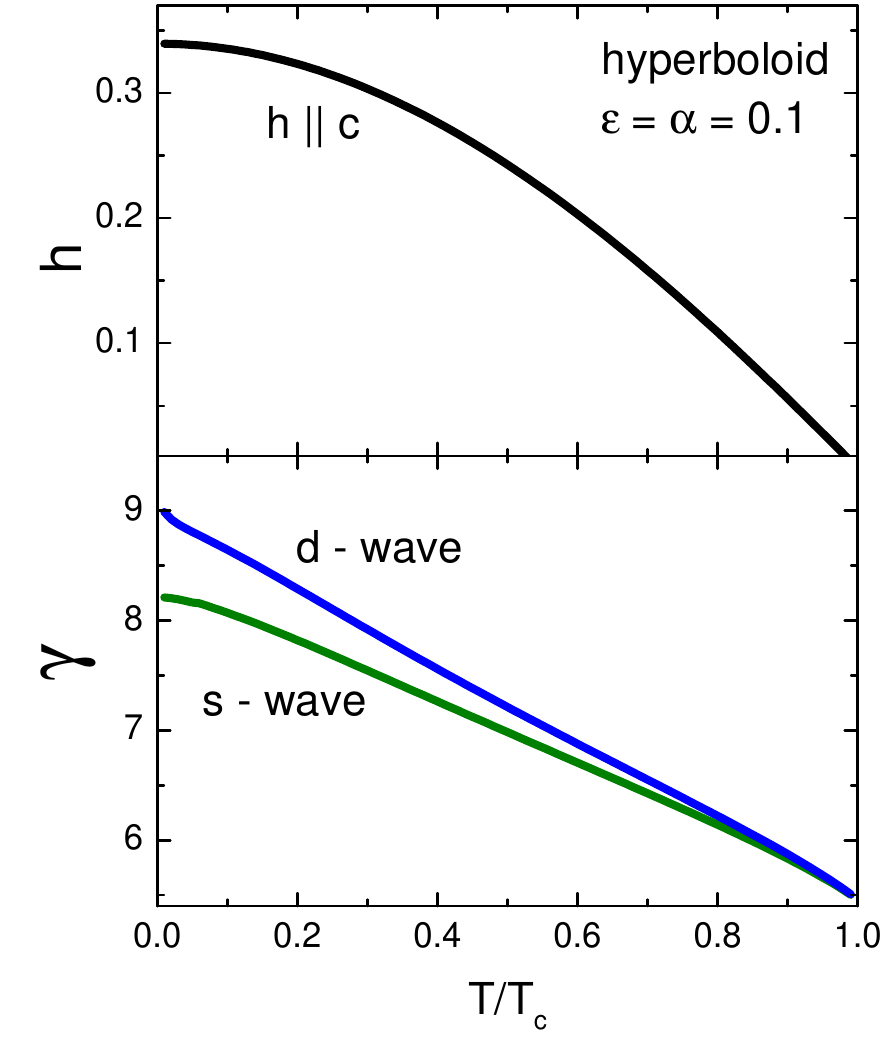}
 \caption{(Color online)  The upper panel:  $h_c(t)$ for Fermi hyperboloid  with $\epsilon=\alpha=0.1$. The lower panel: the anisotropy  $\gamma(t)$ for the same parameters of the Fermi hyperboloid.  }
 \label{fig11}
 \end{figure}

The density of states $N(0)$ is defined by the integral of Eq.\,(\ref{eqDOS}),
which can be written as an integral  over the
solid angle $d\Omega=\sin\theta\;d\theta\, d\phi$:
\begin{equation}
N(0)=
\dfrac{m^2_{ab} v_{ab}}{2\pi^2\hbar^3}
\int\dfrac{d\Omega}{4\pi \sqrt{\Gamma_2(\theta)\Gamma_1(\theta)}}\,,
\label{N(0)1}
\end{equation}
where the integration over $\theta$ is extended from $\theta_m$ to $\pi -\theta_m$.
The Fermi surface average of a function $A( \theta,\phi)$   is
 \begin{eqnarray}
\langle A \rangle
 &=&\frac{1}{D}
 \displaystyle\int\dfrac{d\Omega\,A( \theta,\phi)}{4\pi  \sqrt{\Gamma_2(\theta)\Gamma_1(\theta)}}\,,\qquad\qquad\label{<A>}\\
D &=& \int\dfrac{d\Omega}
{ 4\pi \sqrt{\Gamma_2(\theta,\epsilon)\Gamma_1(\theta,\epsilon)}} \,.
\label{aver}
\end{eqnarray}

For $A$ depending only on $\theta$, one can employ   $u=\cos \theta$:
\begin{eqnarray}
\langle A \rangle
=\frac{1}{D(\epsilon)}
 \int_0^{u_m}\frac{du\,A(u)}{ \sqrt{\Gamma_2(u,\epsilon)\Gamma_1(u,\epsilon)}}\,,\qquad\qquad\label{<A1>}\\
\Gamma_2 =  1-(\epsilon+1)u^2\,,\qquad \Gamma_1  = 1+(\epsilon^2-1)u^2 \,.\qquad
\label{aver1}
\end{eqnarray}
where the upper limit is $u_m=\cos \theta_m=1/\sqrt{1+\epsilon+\alpha}$.
In particular, one obtains:
\begin{eqnarray}
D(\epsilon,\alpha) =  \frac{F(\tan^{-1}\sqrt{(1+\epsilon)/\alpha},1-\epsilon)}{\sqrt{1+\epsilon}}  \qquad
\label{average}
\end{eqnarray}
where $F $ is an Incomplete Elliptic Integral of the first kind.

As for ellipsoids, the relation between $v_{ab}$ and $v_0$ defined in Eq.\,(\ref{v0}) for a one-band situation holds:
\begin{eqnarray}
v_{ab}^3= D \, v_0^3 \,,
 \label{v0-vpar1}
\end{eqnarray}
however, with a different $D$.

For $\bm H\parallel \hat c$, the relevant electron orbits are circular, as for spheres and rotational ellipsoids, and we do not expect  qualitative deviations from the latter. Fig.\ref{hc-hyperb} shows $h_c(t)$ calculated with the help of Eqs.\,(\ref{eq-hc}) and (\ref{mu_c}) for both s- and d-wave order parameters. We note that the anisotropy parameter shown in this figure is suppressed substantially on warming.

Numerical evaluation of the anisotropy  shows that $\gamma(t)$ decreases   on warming,  Fig.\,\ref{fig11}. Reduction of the anisotropy on warming is substantial for a single band open Fermi surface, an interesting observation  given the common belief that the  temperature dependence of anisotropy is a multi-band property. We also note that this reduction is stronger than that of open Fermi surfaces.


\begin{references}
\bibitem{remark1}It is instructive to note that the zero-$T$ value of the ratio of coordinate dependent order parameters at the upper critical field $ H_{c2}(T=0)$ is the same as at zero-field and $T=T_c$.\cite{gamma-model,KS}
\end{references}
  \end{document}